# Bootstrap Cross-validation Improves Model Selection in Pharmacometrics

James Stephens Cavenaugh

ORCID: 0000-0002-7072-2418

Affiliation: Department of Pharmaceutical Sciences, School of Pharmacy and Pharmaceutical Sciences, University at Buffalo, Buffalo, NY, USA

jscavenaugh@gmail.com

1 (505) 780-0030

## ABSTRACT

Cross-validation assesses the predictive ability of a model, allowing one to rank models accordingly. Although the nonparametric bootstrap is almost always used to assess the variability of a parameter, it can be used as the basis for cross-validation if one keeps track of which items were not selected in a given bootstrap iteration. The items which were selected constitute the training data and the omitted items constitute the testing data. This bootstrap cross-validation (BS-CV) allows model selection to be made on the basis of predictive ability by comparing the median values of ensembles of summary statistics of testing data. BS-CV is herein demonstrated using several summary statistics, including a new one termed the simple metric for prediction quality (SMPQ), and using the warfarin data included in the Monolix distribution with 13 pharmacokinetics (PK) models and 12 pharmacodynamics (PD) models. Of note the two best PK models by AIC had the worst predictive ability, underscoring the danger of using single realizations of a random variable (such as AIC) as the basis for model selection. Using these data BS-CV was able to discriminate between similar indirect response models (inhibition of input versus stimulation of output). This could be useful in situations in which the mechanism of action is unknown (unlike warfarin).

# Introduction

Modelers may take it for granted that modeling has value, but non-modeling scientists do not always see the value of modeling the same way modelers do. I recall a conversation with a non-modeling faculty immunologist who commented along these lines: "Modelers cherry pick from the vast array of known facts a subset to include in their models, so what value do they bring? What have we learned that wasn't already known?" For her, most of the value of modeling was probably educational but only in a minor way: a model could clarify current thinking on a topic but could not offer much more. Real discoveries were made by experimentalists like her.

In a sense, she's right, but only if the issue of model selection is ignored, as it often is. From where does the value of modeling come? This is clearly a philosophical question that may have more than one valid answer, but I contend that most of the value of modeling arises from model selection. (I use the phrase *model selection* in a broad sense, involving discrimination of any aspect of a mathematical or statistical model.) Picking model A over model B, when a priori both seem sensible, involves carefully weighing the merits of the models and comparing against known facts. The process of model development, which inherently involves numerous successive instances of model selection decisions along the way, forces clarity of thought and in so doing adds cognitive value. If only one model is presented, then for all practical purposes it's the correct model *by assumption*. The deliberations involved in discriminating between models is the basis for much, if not most, of the cognitive value of modeling.

Model selection, I believe, often seems to be under-appreciated, as if it were a mature science which scarcely needed improvement. The model selection literature is vastly smaller than the corpus of statistical literature not aimed at model selection (as perusing just about any statistical journal will confirm). It's easier to find modeling papers that ignore this issue entirely than it is to find modeling papers that provide an appendix or online supplementary material that goes into any depth to justify the final model chosen. R. A. Fisher, who formalized the notion of likelihood, considered model selection to be outside the realm of theoretical statistics—something for the practical data analyst to take on a case by case basis [1]. Likelihood calculations presuppose some particular model; the model is a given.

So how is model selection done? There are at least four considerations in current practice. The first quite rightly involves subject matter understanding. For example, in a pharmacometrics context a polynomial could be used as a basis for a structural model, but that would have no physiological meaning at all, whereas a compartment model using a sum of exponential terms at least has a very crude basis in physiological reality: drugs permeate some tissues faster than others. The second pillar of current practice involves visualization, beginning with simple plots of one measured quantity against another and moving up to sophisticated graphical diagnostics [2-4]. The third pillar of current practice is based on some sort of statistical hypothesis testing, typically rooted in *P*-values. The point here is that an arbitrary cutoff (such as $\alpha = 0.05$) is used to pick one model over another. The fourth pillar of current practice is rooted in likelihood theory and information theory: the use of a penalized likelihood such as Akaike's information criterion (AIC) or Schwarz' Bayesian information criterion (BIC) or modifications thereof [5, 6]. This is similar in spirit to the third pillar in that the choice of model relies on a single statistic, but instead of using a cutoff the criterion is to take the model with the best (lowest) penalized likelihood score. There are certainly other approaches as well, such as using the minimum description length principle as a desideratum [7] or based on the bootstrap [8], but those approaches tend to be outside mainstream model development, at least as practiced routinely in pharmacometrics. It behooves us to re-examine some fundamental facts about current practice and ask if there is a discrepancy between the way things are done and the way they ought to be done.

Model development in pharmacometrics is an iterative process in which one starts with a simple model (after looking at the data and verifying that they are satisfactory) and then successively considers modifications in a stepwise process known as forward-selection, backward elimination [2, 3]. At each step of the way, whether one is modifying the current working model by the addition of some covariate relationship or another structural term, or whether one is considering deleting an existing part of the current working model and seeing if such a deletion is significant, several graphical diagnostics can be used to help the decision, but often a cutoff is used based on the change in objective function value, $\Delta$ OFV. (The OFV is $-2 \times$ the log of the likelihood; the actual calculation of an approximation to the likelihood depends on algorithmic choices that involve a tradeoff between accuracy and calculation time. Since the likelihood is a product of probabilities (p. 104 of [9]), each of which is between 0 and 1, the log likelihood is negative, and so the OFV is a positive number where the smaller it is, the better the fit.) If the addition of a parameter results in a $\Delta|\text{OFV}| \leq$ *cutoff*, then this parameter is considered insignificant and would not be included on purely statistical grounds (although perhaps the modeler has other reasons for wanting it to be included anyway), and conversely, a $\Delta|\text{OFV}| >$ *cutoff* would result in the model addition being retained. If one formally compares nested models using a likelihood ratio test, the $\Delta$ OFV is approximately distributed according to a $X^2$ distribution for a pre-specified α value and number of degrees of freedom equal to the number of additional parameters estimated in the larger model. For such a distribution with one degree of freedom a difference more than *cutoff* = 3.84 is significant at the conventional α = 0.05 level used in hypothesis testing [2]. For backward elimination a more stringent α level is often used, e.g. α = 0.001. If the models are not nested, then a likelihood ratio test cannot be done, but nevertheless the same cutoff value can still be used as a criterion for retaining the model modification or not.

Several features of this forward addition, backward elimination paradigm should be noted. First, it is by far the dominant paradigm for model selection in pharmacometrics. Second, it is time-consuming. The number of potential models grows combinatorially with the number of model components. With tight deadlines, a modeler might of necessity skip the evaluation of some models that he'd rather look at and settle for what he hopes is good enough. This disadvantage can be offset by the third point: since it is essentially algorithmic, it is amenable to automation. For example, Marc Lavielle developed an automated model building platform for Rsmlx [10]. Fourth, the forward addition, backward elimination modeling approach winds through the sample space, eventually settling on some model as best, but if a different trajectory in the sample space had been followed, perhaps if covariate B had been evaluated prior to covariate A, then a different final model may have been selected. This is an example of what Leo Brieman called the "Rashomon effect" [11]: in complicated multivariate data sets, the same data can often be explained almost as well by a large number of very different models. It may be in vain to try to find a global optimum when there are many predictors, as the differences in objective function value may be insignificant. While the forward addition, backward elimination procedure may fail to give the best model for a given data set, fortunately the best model it does find might not be too bad. A few other fundamental facts pertinent to this procedure are worth strongly emphasizing (in order of increasing concern):

1. The justification of the cutoff is based on *P*-values that are only approximately valid.

2. Each model choice is based on a single realization of a random variable.

3. It aims to improve fit, not prediction.

Each of these points will be more fully discussed further below.

**The justification of the cutoff is based on *P*-values that pertain to a distribution that is only approximately valid.** The basis for favoring one model over another—the $\Delta$ OFV—is only asymptotically $X^2$ distributed, which implies that *P*-values for the $X^2$ distribution are only approximately valid; they are not exactly valid for the real situation at hand, so the statistical justification of changes to the working model actually being made with a $(1 - α)\%$ probability is incorrect. This weakens the justification for using a particular $\Delta$ OFV as a cutoff for accepting a model change (such as 3.84 for the addition of a single parameter at α = 0.05). It might be true that Δ OFV sufficiently lies close to a $X^2$ distribution [12], but there is no guarantee. The quality of the approximations depends in part on the accuracy used for the likelihood calculation (and in part on the fact that sampling distributions of random variables are relevant here, discussed next). Better results are obtained using more accurate likelihood approximations [2, 13-15].

**At each decision point in the model development process, the model choice depends on a single realization of a random variable.** That the *P*-values pertain to a distribution that's only approximate is one issue, but the fact that they are used at all to provide model justification is another issue. *P*-values have long been controversial, as evidenced by the many authors and papers who have criticized *P*-values for decades. For a good place to start, see the ASA's statement on *P*-values and the papers and reviews therein [16]. Most of these deal with the fact that *P*-values are easily and often misunderstood, but misunderstanding is a problem of statistical education, not the statistics themselves. Although it is not widely appreciated, *P*-values are random variables with their own probability distributions—just like the sample mean or the range. (Here for clarity I follow the statistical convention of using a capital letter to denote a random variable *X* and some particular realization of it by a lower case letter, *x*. Hence *P*-values emphasizes the fact that these quantities are random variables pertinent to some probability function, stopping rule, and sample space, and *p*-values emphasizes a particular realization based on a particular data set. Unfortunately statisticians rarely follow this convention for *P*-values.) When a particular *p*-value is obtained in order to make an inference, that *p*-value is almost always obtained using the full, actual data set. Some statistics, like the median, are robust to sampling variation, whereas other statistics, like the range, are not at all robust. As a basis for making inferences intended to be reflective of an unobserved, underlying population from which the observed sample is merely an instantiation, it would be very desirable if *P*-values were robust to sampling variation. How robust are they? Despite being a mainstay of routine statistical inference for decades, this question was largely ignored. It was finally addressed in 2011 by Dennis Boos and Leonard Stefanski [17] who noted:

> Good statistical practice demands reporting some measure of variability or reliability for important statistical estimates. For example, if a population mean $\mu$ is estimated by a sample mean $\bar{Y}$ using an independent and identically distributed (iid) sample $Y_1, \ldots, Y_n$, common practice is to report a standard error $s/\sqrt{n}$ or a confidence interval for $\mu$, or their Bayesian equivalents. However, the variability of a *p*-value is typically not assessed or reported in routine data analysis, even when it is the primary statistical estimate.
>
> More generally, statisticians are quick to emphasize that statistics have sampling distributions and that it is important to interpret those statistics in view of their sampling variation. However, the caveat about interpretation is often overlooked when the statistic is a *p*-value. With any statistic, ignoring variability can have undesirable consequences. In the case of the *p*-value, the main problem is that too much stock may be placed in a finding that is deemed statistically significant by a *p*-value < 0.05. In this article we systematically study *p*-value variability with an eye toward developing a better appreciation of its magnitude and potential impacts, especially those related to the profusion of scientific results that fail to reproduce upon replication.

Their paper is certainly worth reading, but the short answer is that *P*-values have lots of variability and are basically only valid to an order of magnitude. It is pointless to get excited if $p = 0.04$ and disappointed if $p = 0.06$ for a conventional $\alpha = 0.05$. As Boos and Stefanski showed, if you tweak the data slightly, for example by the bootstrap, you'll get a different *p*-value—possibly very different. In my view, this fact is very damning for the use of individual *p*-values as a primary inferential tool.

Although *P*-values and hypothesis testing are by far the most common model selection tools in science as a whole, in pharmacometrics model development one looks at the Δ OFV or a penalized likelihood such as AIC rather than actually at a *p*-value. As with *P*-values, the log-likelihood (and hence OFV) and penalized likelihoods are random variables with their sampling distributions, depending only on the model and the particular realization of data obtained in nature from some unknown distribution. The difference between two OFVs is also a random variable because the two OFVs depend on different models but the same data. Boos' and Stefanski's paper investigated the robustness of *P*-values, not AIC or BIC scores; however, AIC and *P*-values (based on the *F*-test) are related for nested models so that if you know one, you can get the other (p. 27 in [3]). Hence it would not be surprising to find that AIC and the like are also not very robust. The Results section of the present work bears out this hunch. Put simply, it can be dangerous to put much faith in a single AIC value or Δ OFV.

**The forward addition, backward elimination procedure aims to improve a model's fit to current data, not its predictive ability.** After the initial graphical exploration of the available data, all subsequent stages of the forward addition, backward elimination paradigm are aimed at improving the fit between the working model and the observed data. In discussing the model development process, the NONMEM V manual [18] states, "A more complex model is acceptable only if the complexity can be justified by some significant improvement in the fit. To evaluate whether this has been accomplished, several measures should be examined; no single measure suffices." The implicit assumption is this: *the better the fit, the better the model*. This assumption should be re-considered, since not surprisingly working toward improving the fit can lead to overfitting. Overfitting is undesirable because then one is fitting not only the signal (the basic patterns in the data) but also the random noise, and as a result the predictive ability of the model worsens. The usefulness of a model depends on its intended purpose, but generally speaking a model that makes good *predictions* of future data will be far more useful than a model that merely fits observed data well.

It is with these shortcomings of the current forward-addition, backward elimination paradigm in mind that the present work was undertaken.

# Theoretical

## Bootstrap cross-validation

In Bayesian inference one updates one's prior beliefs with observed data according to Bayes' rule. Basing inferences on data that were actually obtained is obviously appealing, but it comes at the cost of requiring a prior, which, rightly or wrongly, has been controversial. In contrast, frequentist inference is based on sampling distributions. In other words, frequentist inferences depend not on only what was observed but also on data sets that could have been observed but weren't [19]. In this sense resampling approaches such as the bootstrap and cross-validation are purely frequentist: they generate many sets of "fake data" that are not identical to the actual observed data set but which are derived from the observed data and hence are very plausible.

Nonparametric bootstrapping is resampling with replacement so that from a single, original sample, one gets a set of bootstrap samples; their distribution mimics the unobserved, underlying distribution from which the original sample was drawn. The emphasis is almost always to get an estimate of the variation of some statistic of interest, often one that would be difficult to do from purely theoretical grounds. For hierarchical or mixed effects models, as population PK and PD models typically are, the resampling should be at the level of the individual. In cross-validation the data are segregated into mutually exclusive subsets, one for "training", i.e. for ordinary parameter estimation as one would do with the entire data set, and the rest of the data is used for testing the results from the training set. The goal is to estimate how well a model fits data that were not used to develop the model, i.e. get an assessment of prediction error. While external validation using an independent data set is most persuasive for predictions, cross-validation is the next best thing to more real data, which are usually financially or otherwise impractical to obtain.

Although the emphasis in the bootstrap is typically parameter estimation and related confidence intervals, one can also use it as a basis for cross-validation. To do so one needs to keep track of not only which items were selected by the bootstrap's sampling with replacement, but also which items were not so selected. These constitute the training and testing data sets respectively. The probability that any one item will be selected by the $b$th bootstrap sampling procedure is $P$ (observation $i \in$ bootstrap sample $b$) = $1 - (1 - 1/N)^N = 1 - e^{-1} \approx 0.63212$ (as $N \to \infty$), and hence this is known as the 0.632 bootstrap [19, 20]. However, I shall refer to it as bootstrap cross-validation (BS-CV), which I think is more descriptive. The probability of not being selected is $1 - (1 - e^{-1}) \approx 0.36788$. Roughly 2/3 of the items will be selected (training set) and 1/3 will be in the testing set. In this sense the bootstrap is a bargain: you get cross-validation almost for free. (That said, the R code for doing a bootstrap (sampling with replacement) is just 1 line, but actually implementing this for mixed effects models is much more involved. The R functions are included in the supplemental Appendix.)

The bootstrap cross-validation work flow goes like this: data → training and testing subsamples → do some statistical procedure(s) → summary stats on both training and testing data → repeat → get ensemble of testing and training errors → report such errors. From a frequentist perspective this is a much better measure of a model's true accuracy than drawing all inferences of accuracy assessment from only 1 sample. For model selection, pick the model with smaller testing errors (and hopefully smaller variance). The notion "the better the model fits the data, the better the model" is naive because it ignores the problem of over-fitting. However, for the ***testing*** data used to evaluate predictive ability, it is true: the better the fit, the better the prediction, and hence the better the model.

## Summary statistics for use with BS-CV

The fundamental hypothesis of this research is this: *Can the 0.632 bootstrap be applied to nonlinear mixed effects models such as those arising in population PK and PD so as to provide a useful tool for model selection?* The short answer is yes, as shown in the Results, but to do so requires some summary sample statistic to serve as a common basis for comparison across models. Many summary statistics are possible, and I wanted to cast a reasonably wide net. Hence a secondary question is this: *Which of the following summary statistics is most useful for model selection?: –2LL, AIC, BIC, RSS, RMSE, SAD, MAD, SMPQ, ε-shrinkage*. These are defined below:

- –2LL is –2 × log(likelihood), the likelihood being the primary probabilistic measure of fit. Same as OFV.
- AIC $\equiv$ –2LL + 2$p$ is Akaike's information criterion. Here $p$ is the number of parameters.
- BIC $\equiv$ –2LL + log($N$) $p$ is Schwarz' information criterion, a.k.a. Bayesian information criterion.
- RSS $\equiv \Sigma (Y_i - \hat{Y})^2$ is the residual sum of squares.
- RMSE $\equiv \sqrt{[(1/n)\Sigma (Y_i - \hat{Y})^2]}$ is the root mean squared error.
- SAD $\equiv \Sigma |Y_i - \hat{Y}|$ is the sum of absolute deviations.
- MAD is the median absolute deviation.
- SMPQ $\equiv$ –log(1 – $r^2$) is the simple metric for prediction quality. Here the sample correlation coefficient $r$ pertains to the zero-intercept goodness of fit line of individual predictions versus observations.
- ε-Shrinkage $\equiv$ 1 – SD(IWRES) where SD(.) denotes standard deviation and IWRES are the individual weighted residuals [21].

The above summary measures include both probabilistic (–2LL, AIC, BIC) and non-probabilistic, purely geometric measures of fit that depend only on the residuals: RSS, RMSE, SAD, MAD, SMPQ. The ε-shrinkage is somewhat in between, in that strictly it depends only on the residuals, but one would not normally calculate IWRES in the absence of a probabilistic context. The weighting factor by which the residuals are weighted in IWRES does not depend on the residuals.

RSS and RMSE depend on squared error, as is common throughout statistics, so extreme values will count more. In the analogous quantities SAD and MAD all errors count the same, i.e. linearly with magnitude. It is not easy to see which would be more useful for model selection a priori.

The usual Pearson correlation coefficient $r$ is a measure of linear association between two random variables and not a measure of fit (p. 18 in [3]). However, here the line of interest is the goodness of fit line of identity of observations versus predictions, so the particular $r^2$ here *is* in fact an excellent measure of fit. It is more convenient to modify it as seen here, since $-\log(1 - r^2) \to \infty$ as $\Delta(\text{observed, predicted}) \to 0$. Hence this statistic is indeed a simple metric of prediction quality, easily obtained from summary statistical output of many statistical programs. The predicted values refer to individual predictions (IPRED in NONMEM jargon) as opposed to population predictions (PRED in NONMEM). The term "simple" in "simple metric for prediction quality" is meant to imply that other similar quantities could also be made. None of them are as intuitively desirable in my opinion as the SMPQ, although this could be explored in future work.

As data become sparse empirical Bayes estimates (EBEs) approach their population distribution ("η-shrinkage") [22], and also the distribution of individual weighted residuals (IWRES) shrinks towards 0 ("ε-shrinkage") [21]. This is therefore a measure of under- versus over-fitting. NONMEM calculates ε-shrinkage, but Monolix does not. However, Monolix does provide two estimates that one could use to calculate ε-shrinkage, one being from EBEs and the other from simulations. Both of these variations were used herein.

# Methods

For pharmacometric software to execute BS-CV it is necessary to be able to get likelihood values (or at the very least residuals) without doing any optimization; this is for the testing data. If one uses NONMEM one can use the `MAXEVAL=0` option on the estimation step. Monolix (used herein) uses the SAEM algorithm so that in principle the estimation task is needed. However, it is possible to fix parameters to set the number of iterations to 1 and then use the R API function `getEstimatedLogLikelihood()` with either the default importance sampling (more accurate) or the linearization approximation (faster). Details are described further on the Lixoft website [23] and in Chapter 9 of [9].

To execute BS-CV, $B = 100$ bootstrap iterations were generated according to code in the Appendix from the warfarin data [24] supplied with Monolix. These data describe the results of a single oral dose of warfarin (1.5 mg/kg body weight) to 32 subjects (27 males, 5 females) at $t = 0$. Some subjects' first measurement was at 24 h, too late for good information on the absorption phase. The PK variable (y1) is concentration and the PD (y2) is effect (anti-coagulant; prothrombin complex response). Age, weight, sex were covariates in the data set, but since there were so few females, sex was set to IGNORE in Monolix. These BS data were saved to disk. The same data were used for all models, hence making a fair comparison across models. A series of PK and PD models was generated using Monolix 2018R2 [25], and then an R script incorporating the Monolix R API was used to get quantities of interest and save them to a data frame. This data frame was then used to make appropriate graphs using the R package ggplot2 [26]. Computational details are in the supplemental Appendix.

## Pharmacokinetic Modeling

The 13 pharmacokinetic models tested are summarized in Table 1. Following Monolix' terminology, the Observation Model refers to the error model, and the Individual Model column shows the form which the inter-individual variability takes for each parameter, including covariates if any. Models 1-6 are all 1-compartment models. Model 1 is just the basic 1-compartment model with no covariates. All the other models have a lag time ($t_{lag}$ parameter). Models 7-13 are 2-compartment. The structural models are as follows [27] (where $C(t) = 0$ for $t$ $< t_{lag}$, $D$ is dose, and bioavailability is assumed 1): $C(t) = \frac{D k_a}{V_d (k_a - k_e)} \left( e^{-k_e(t - t_{lag})} - e^{-k_a(t - t_{lag})} \right)$ for models 1-6 (but $t_{lag} = 0$ for model 1) and $C(t) = D \left( A e^{-\alpha(t - t_{lag})} + B e^{-\beta(t - t_{lag})} - (A + B) e^{-k_a(t - t_{lag})} \right)$ where

$$\alpha = \frac{\frac{Q}{V_2} \frac{Cl}{V_1}}{\beta}$$

$$\beta = \frac{1}{2} \left( \frac{Q}{V_1} + \frac{Q}{V_2} + \frac{Cl}{V_1} - \sqrt{\left( \frac{Q}{V_1} + \frac{Q}{V_2} + \frac{Cl}{V_1} \right)^2 - 4 \frac{Q}{V_2} \frac{Cl}{V_1}} \right)$$

$$A = \frac{k_a}{V_1} \frac{\frac{Q}{V_2} - \alpha}{(k_a - \alpha)(\beta - \alpha)}$$

$$B = \frac{k_a}{V_1} \frac{\frac{Q}{V_2} - \beta}{(k_a - \beta)(\alpha - \beta)}$$

The parameters $Cl$, $V_1$, $V_2$, $V_d$, and $Q$ have their usual PK meanings.

An earlier model selection project executed using NONMEM, which allows for more flexible error modeling than does Monolix, showed that manipulating the error model gave a considerably better fit at the cost of only one additional parameter. This was the motivation for using 3 error parameters for models 7 and 8, whereas all the other models have 2.

## Pharmacodynamic Modeling

Pharmacokinetic model 9 was deemed to be the best of the PK models considered (as discussed later), and was therefore used with all individual PK parameters held fixed for all the pharmacodynamic models tested. Besides providing another example of the utility of BS-CV, an additional goal for PD modeling was to see if one could distinguish between indirect response model (IRM) type 1 (inhibition of input) from IRM type 4 (stimulation of output) based purely on statistical grounds. This could be useful in cases where the mechanism of action is unknown, unlike warfarin. These two models have the same general shape but very subtle differences. I investigated 12 PD models in 4 patterns of 3 each: first the basic model, then with a Hill parameter, then with the most suitable covariates identified by a plot of parameter correlations versus covariates (using the Hill parameters). Models 1-3 are variations on IRM-1; models 4-6 are variations on IRM-4; models 7-9 are also variations on IRM-1 but with Imax = 1 (fixed); and models 10-12 are combined IRM-1 and IRM-4: they have both inhibition of input and stimulation of output. With the covariates used, the list of PD models is as follows:

Model 1: IRM 1 (inhibition of input); 4 parameters: R0, kout, Imax, IC50

Model 2: Ibid + Hill parameter (= "gamma")

Model 3: Ibid + covariates: Hill(LnWt70), IC50(LnAge30), kout(LnWt70)

Model 4: IRM 4 (stimulation of output); 4 parameters: R0, kout, Emax, EC50

Model 5: IRM 4 with Hill parameter

Model 6: Ibid + covariates: Emax(LnWt70) and EC50(LnAge30)

Model 7: IRM 1 with Imax := 1.  3 parameters: R0, kout, IC50

Model 8: Ibid + Hill parameter

Model 9: Ibid + covariates: IC50(LnAge30), Hill(LnWt70)

Model 10: IRM 1 combined with IRM 4.  6 parameters: R0, kout, Imax, IC50, Emax, EC50

Model 11: Ibid but with 2 Hill parameters

Model 12: Ibid + covariates: IC50(LnAge30), kout(LnAge30), Hill1(LnWt70)

The error models all used the additive + proportional error model (i.e. combined1 in Monolix' terminology). Similarly to the PK models, all the parameter distributions of the structural models were lognormally distributed. Just to be clear on notation, "Hill(LnWt70)" means that the individual Hill parameter is adjusted by weight, log-normally scaled by the mean weight (70 kg): $\log(H_i) = \log(H_{\text{pop}}) + \beta_H \log(wt_i/70) + \eta_{H,i}$ and similarly for the other covariate relationships.

# Results

## PK results

The top, middle, and bottom panels of Figures 1-10 show the various summary statistics of interest described above for models 1-13 for the original data, the bootstrap training data, and the testing data respectively.  For most of the figures the three panels are on the same scale.  Note that for the statistics –2LL, AIC, BIC, RSS, and SAD, the median values for the training data are generally very different from the medians for the testing data, reflecting the fact that in the testing data set there is simply less data (due to smaller sample sizes) and less information.  In contrast the statistics RMSE, MAD, SMPQ and ε-shrinkage are not so sensitive to the sample size and information content.

The likelihood based figures (Figures 1-3) are all very similar.  These summary statistics for the original data are slightly higher than the medians of the corresponding statistics of the training data, suggesting that there may be slightly more information in the original data than in a typical bootstrap sample.  As one would expect, Figure 1 shows a steep drop in –2LL (better fit) when going from model 1 to model 2, i.e. the inclusion of a lag time for absorption.  The addition of covariates further improves the fit, but less so, and the same is true for going from a 1-compartment model to a 2-compartment model.  Figures 2 and 3 show the analogous AIC and BIC plots respectively.  There is nothing unusual in either the –2LL, AIC, or BIC plots with respect to the original or the training data for any of the models.  Based on the traditional criterion of choosing the model with the lowest AIC, the best model would be model 8 (AIC = 660.5397), followed closely by model 7 (AIC = 660.7185). Recall that these two models have three error parameters, whereas all the others have two.  In contrast, the testing data show a much worse fit (higher –2LL, AIC, and BIC) for models 7 and 8 than for the other 2-compartment models (going by medians for each model).  Model 7 is even worse than the basic, no-frills model 1.  Model 5 is the best (slightly) among the 1-compartment models, and the other 2-compartment models are all about the same.

Figure 4 shows that the residual sums of squares for the original data are all similar regardless of model, and in fact models 7 and 8 have among the highest RSS.  However, in the training data models 7 and 8 show dramatically lower RSS, indicating a much better fit.  This is overfitting as seen very clearly in the testing data: models 7 and 8 have by far the worst predicted fits.  Model 7 is worse than model 8.  Figure 5 tells the same story for the RMSE, which is to be expected given the similar nature of the statistics.

Figures 6 (SAD) and 7 (MAD) are also similar to Figures 4 and 5. It is a bit easier to discriminate among the other models using Figure 6 than 4 since the former does not change the ordinate scale as drastically, dealing with the magnitudes of the residuals rather than the magnitudes of the squares of the residuals. Models 7 and 8 are shown to be by far the worst with these statistics (highest predicted errors), but surprisingly model 1 was the best among the 1-compartment models, as judged by SAD for both the training and the testing data sets, and also by the original data. This contrast between the likelihood (–2LL, AIC, BIC) and the pure residual (SAD, MAD) results (apart from models 7 and 8) is a bit surprising and difficult to explain. The other 2-compartment models are all very similar. The results for MAD (Figure 7) are very similar to those of SAD (Figure 6).

Figure 8 shows SMPQ, which is arguably the best statistic in the set of statistics considered on a priori grounds. It is a direct measure of how well a model's predictions fit the observations. The original data and the training data show that the basic model, model 1, has the worst fit, models 2-5 are much better and are all about the same, and the 2-compartment models are better still and also all about the same. With so few covariates this warfarin data set is a very simple, so it is not surprising that models differing only by the arrangement of two error parameters (combined1 versus combined2) behave so similarly. The testing data are similar except for models 7 and 8, which show a remarkably worse fit. Model 9 has the highest median score (barely) for the testing data, and therefore that was the model that was used going forward with the PD modeling.

Figures 9 and 10 show ε-shrinkage calculated using either the EBEs or simulated data. Unlike with most of these plots, the scales for the original data were left to their default settings to highlight the differences, but they are similar to the medians for the training data. The training data ε-shrinkage using the EBE-derived values are small but positive (Fig. 9), whereas the values using simulated data are very close to 0 except for models 7 and 8 (Fig. 10). The testing data in Fig. 9 had medians all near 0 (some above, some below), except for models 7 and 8, while the testing data in Fig. 10 were all negative. In both figures the training data show strongly positive ε-shrinkage for models 7 and 8, but the testing data show strongly negative ε-shrinkage for these same models. It is not obvious why this should be so. The medians for all the testing data are negative for ε-shrinkage via simulation. The ε-shrinkage values are intriguing but only significant for models 7 and 8, which have different signs between the training and testing data. The ε-shrinkage by simulation (sampling from the conditional distribution) ought to be more accurate than by EBEs for the same reasons that sampling from the conditional distribution is preferred in Monolix, as a way to mitigate the effects of shrinkage [28].

## PD results

As to whether one can distinguish between IRM-1 and IRM-4 based on purely statistical grounds, the short answer is yes. Model 1 has a much better fit based on likelihood measures than does model 4, but not based on other measures such as RSS or SAD. In this sense the PD data do not tell as clean a story as do the PK data. If one includes additional parameters such as the Hill parameter and covariates, even this distinction is eroded.

For the warfarin PD data used, indirect response model 4 (stimulation of output) by itself, with no additional parameters, was noticeably worse than the other models, such as IRM-1 (inhibition of input) as seen by –2LL, AIC, and BIC even using the original data. Figures 11 (–2LL), 12 (AIC), and 13 (BIC) are all very similar to each other and showed similar patterns for training and for testing data. A large improvement to fit was seen for IRM-4 when adding a Hill parameter (model 5 compared to model 4), but small improvements to fit were seen as one added parameters to the other models.

Unlike the likelihood metrics, the testing data panels in Figures 14 (RSS) and 15 (RMSE) show evidence of overfitting in going from model 2 to model 3 and again from model 11 to model 12. These changes involve the addition of covariates to the previous model. Models 1, 2, and 6 are the best by RMSE.

Figures 16 (SAD) and 17 (MAD) tell the same story as do Figures 14 and 15, but the differences are less pronounced for SAD than for RSS. At least for these data and models, MAD seems to be more useful than SAD.

Clearly the addition of covariates does not help the predictive power for the PD models considered for this data set based on these metrics. (Compare models 2 and 3; 5 and 6; 8 and 9; and 11 and 12.)

Figure 18 (SMPQ, i.e. $-\log(1 - r^2)$) confirms that the addition of covariates isn't helping, except when going from model 5 to model 6 and maybe from model 8 to model 9. If one were to focus only on improved fit, one would not pick model 1, but model 1 has the highest median SMPQ score for testing data among the candidate models and hence the best predictive ability (however, the values for all PD models are extremely similar as shown Table 2). These data are another example of how pursuing the best fit with a model, such as in the usual stepwise regression approach used ubiquitously in pharmacometrics, would lead to a model that is not the best in predictive ability.

Figure 19 shows $\varepsilon$-shrinkage based on EBEs, and those medians are all positive. Figure 20 shows the analogous $\varepsilon$-shrinkage based on simulations, and those medians are all negative. The significance of the sign is unclear to me at this time. The relative values of the medians across models in both graphs is suggestive of the same patterns seen in the other graphs.

Individual plots are contained in the Supplemental Information.

## Discussion

With respect to coaxing meaning out of data, the question, "*What do I find persuasive?*" involves both one's modeling philosophy and human psychology. The philosophical foundations of statistics lurk just beneath the surface and inform and guide the actual calculations (Bayesian, frequentist, and so on). Psychological considerations such as familiarity, consistency with prevailing theories, whether one follows historical norms, rhetorical ability, and the reputation of the analyst are also important—especially with regards to persuading others. Often modelers focus their efforts on developing models which fit the available data very well. This is the guiding philosophy underlying the forward selection, backward elimination approach that is ubiquitous in pharmacometrics. The problem with "the better the fit, the better the model" is that working to improve the fit can lead to over-fitting. A better guiding philosophy would be to focus on making models which have good predictive ability and are parsimonious in their explanatory ability.

The most persuasive, highest impact models make surprising predictions that are externally validated in subsequent experiments, such as the Standard Model in particle physics. However, requiring additional data sets from subsequent experiments is a very high bar to persuasion and has its own set of problems, such as cost, time, differences in experimental setup, and the subtle differences in skills of the people running the experiments. Cross-validation offers a computationally demanding but otherwise cost effective approach for internal validation, allowing models to be compared on an even footing based on their predictive ability.

There are multiple forms of cross-validation. At one extreme would be 2-fold cross-validation, where half of the data are reserved for testing and half for training. Since only half of the data (one could use even less with large data sets) are used for model generation (training), models which really need to have lots of data to support them would be biased against; thus 2-fold CV favors simpler models. At the other extreme would be leave-one-out cross-validation (LOOCV) where only a single datum is omitted from the training set. LOOCV is nearly unbiased but has high variance [20]. Bootstrap cross-validation lies in between these extremes. It has an overall good balance between bias and variance [20, 29]. Although BS-CV has greater bias than LOOCV, it has lower variance and better performance and tighter confidence intervals [30]. Adjustments can be made to lower the bias [31], but they would complicate the discussion and are unnecessary for model selection. By keeping track of which items are not selected in any given bootstrap iteration, this research illustrated the utility of BS-CV as a practical means of model selection in pharmacometrics. Future work should include efforts to quantify the bias, variance trade-off using simulated data sets of varying sizes [12], varying types and numbers of covariates, different dependent variables (continuous, time to event, and so on), and various cross-validation approaches.

As shown in [29] the bias of BS-CV (0.632 bootstrap) can vary significantly depending on the data set, so it is important to use realistic PK or PD data sets to get a richer set of experience in the pharmacometrics context. Simulated data allow one to make an assessment of bias. However, when using simulated data one should remember that although there is a true generative model, for a data set of finite size the best model may not necessarily be the model used for the simulations; one can generate data sets that are better explained by simpler models. This hearkens back to the question of what is persuasive, and in part for this reason a real data set was used for this study.

Besides multiple forms of CV, there are also multiple possible summary statistics one could use as the basis for comparison across models. In this study several probabilistic statistics (likelihood based) and several non-probabilistic statistics (based only on residuals) were examined. When used with testing data the SMPQ provides an especially simple and versatile assessment of predictive ability. One should distinguish between the summary statistics used to provide a basis for a "common currency" across models (i.e., –2LL, AIC, BIC, RSS, RMSE, SAD, MAD, SMPQ, and ε-shrinkage) from the summary statistic actually used to compare them, in this case the median. Other measures of overlap between clouds of points pertaining to BS data of various models could be used, for the purpose of model selection such complications seem unnecessary. Comparing medians for predictive accuracy by BS-CV seems adequate.

Resampling as by BS-CV greatly adds to the time taken for modeling. Does one always need to do some form of cross-validation such as BS-CV? No, only when it's important to have good predictive models (but if it's not important, why is the modeling done?). If the final model will be used as the basis for simulating expensive clinical trials or if important clinical or financial decisions depend on the model, then it absolutely makes sense to get the model selection right and to have as high a confidence in the chosen model as possible. If one were completely confident that one was erring on the side of under-fitting and not over-fitting, then there would be no need to do CV of any sort. This may be true at the beginning of a modeling endeavor, using simple compartmental models and before lots of covariates are included with possibly different functional forms. However, as candidate models become more complicated, the risk of over-fitting grows, and hence the need to evaluate predictive performance and not merely fit. To balance the desideratum of finishing a modeling endeavor in a timely manner with the desideratum of having models with good predictive performance, my recommendation would be to do BS-CV either on only the top 10-20 final candidate models or to do BS-CV on the top 5 or so candidate models as one proceeds along with either manual forward selection, backward elimination or an automated equivalent and to re-do BS-CV from time to time as new models are generated. In order to have a benchmark for models known not to suffer from over-fitting, I suggest including the simplest model examined (such as a simple 1-compartment model with no covariates in the present case) in at least the first round of BS-CV.

An obvious question is how many iterations of BS-CV should be done. To get a very good estimate of the covariance between some random variables, it may be necessary to do very many (> 1000) iterations. However, the issue here is model selection, not getting very tight confidence intervals for some parameter. For BS-CV it is only necessary to do enough bootstrapping that one can have good confidence in comparing the medians (assuming that the median is the basis for comparison). In the present study 100 BS iterations were done, which should be plenty to have enough confidence in the medians to make a comparison. Note that these bootstrapped data sets were saved to disk prior to doing any modeling on them in order to be sure to have a fair comparison from one model to another without the complication of comparing different models with different sets of bootstrapped data. If that were not the case, more iterations would be needed, but how many more is unclear.

The phrase "to have confidence in the medians" may suggest that standard errors of the medians need to be calculated so that formal statistical tests could be done, instead of merely comparing medians directly. If one really wanted to, one could do a bootstrap of the bootstrapped summary statistics and thereby get confidence intervals for the median values for the statistics of interest and even do permutation tests to put a *p*-value on the difference between the medians of two models. (In the spirit of cautioning against drawing inferences using single instances of random variables, I would suggest using a distribution of *p*-values—if that's what one finds persuasive—rather than single instances.) However, if one is tempted to look for *p*-values or for the overlap of confidence intervals for BS-CV summary statistics between models, one really needs to remember that *statistics is quantitative epistemology* and to think carefully about the philosophical underpinnings and ask, "What exactly am I trying to accomplish?" In hypothesis testing one has a model to be nullified with data—the null model. A situation in which one might want to do formal nonparametric hypothesis testing between the distributions of bootstrapped summary statistics would be if you're designing a clinical trial and model A will cost you $$x$ whereas model B will cost you more. Now you do have a reason to prefer model A, and you would want to know if model B really is that much better as to justify model B, so it would make sense to ask (for a one-sided test), "Is model B significantly better than model A?" Using model A as the null hypothesis would be sensible in such a case: if you fail to reject model A as inadequate, then stick with it. However, if the goal is just to pick a model with good predictive ability and there is no a priori bias for or against any one model, then I see no point in trying to put a *p*-value on the summary data just to be able to say, "Models A, B, and so on are significantly different from each other (a phrase which the American Statistical Association finally officially deprecates [32, 33]) according to a traditional hypothesis test, which ignores *P*-value variability anyway [17, 34]." For the task of picking a single model to use for downstream purposes such as clinical trial simulations, comparing medians directly is adequate.

Forward addition, backward elimination is an algorithmic paradigm which yields models that are probably by and large good models (due to the Rashomon effect) but which nevertheless has a shaky epistemological foundation. The results of BS-CV may or may not agree with model selection based on traditional criteria such as AIC. In the present case PK models 7 and 8 (with 3 error parameters, versus 2 for all others) were the best by AIC on the original data but were by far the worst models when seen in the light of BS-CV. By comparing the testing data to the training data across the various summary statistics, a consistent picture emerges in which these models are seen to be highly overfit. Apparently an additional parameter in the error model gives one a greater ability to fit (and over-fit) data than does an additional parameter elsewhere. This example should serve as a warning against using AIC or BIC on the original data as the basis for model selection. For the warfarin PD data used, indirect response model 4 (IRM-4, stimulation of output) by itself, with no additional parameters, was noticeably worse than the other models, such as IRM-1 (inhibition of input) as seen by both AIC using the original data and also by BS-CV testing and training data. A large improvement to fit was seen for IRM-4 when adding a Hill parameter, but small improvements to fit were seen as one added parameters to the other models. The model with the highest SMPQ score was model 1, which also happens to be the accepted model for warfarin based on its mechanism of action [35]. By looking at medians of distributions of random variables instead of single instances, BS-CV gives greater assurance in the results. BS-CV is thus seen to be useful for discriminating between IRM-1 and IRM-4. However, the generality of this should be studied further in other drugs and data sets.

The use of random variables such as *P*-values, AIC, and BIC based only on the original data and a model of interest has been the standard way of frequentist model development throughout most of science, including pharmacometrics. *A single realization of a random variable is implicitly treated as if it had large sample asymptotic generality suitable for making trustworthy decisions regarding models*. Unfortunately this is just not so [17, 36]. AIC has been shown to be asymptotically equivalent to leave-one-out cross-validation [37, 38], but the present work illustrates the danger of relying on single realizations of asymptotically valid statistics. In my opinion *such methods ought to be considered obsolete model selection approaches suitable for the time when they were invented, before powerful computing became available*. Why rely on asymptotics when now we can use powerful computing to do better [19]? Relatively little has been written regarding the variability of penalized likelihoods [39, 40], but this work illustrates the fact that they too are random variables and that perturbing the data, as e.g. by taking bootstrap samples, will result in a distribution of such values. Since the penalty terms in AIC and BIC are intended as valid only in large sample approximations, using a distribution of AIC or BIC values probably offers no advantage over using the distribution of log-likelihood itself. With BS-CV one can directly compare the medians for log-likelihood for testing data for predictive accuracy.

Future work could involve the use of NONMEM and other error models (Monolix presently lacks the same flexibility for error models that NONMEM has), other outcome measures (e.g., survival curves, logistic regression), improved modeling tools which could make it easier to do BS-CV, and comparisons to other methods of cross-validation, including the extremes of 2-fold and LOOCV as noted above. What are the pros and cons of BS-CV compared to these extremes, especially as they relate to the bias vs. variance tradeoff? Are the results one gets sensitive to the algorithm used to calculate the conditional likelihood? Does the ordering of models change when changing algorithms? There are variations on the basic bootstrap (e.g., stratification). Since BS-CV *is* bootstrapping but with the essential distinction of keeping track of which items were not selected in a particular bootstrap iteration so as to have a testing data set, each of the bootstrap variations could also be adapted for BS-CV. To adapt BS-CV for automation such as for genetic algorithms [41, 42] it may be helpful to quantitate overfitting by the degree of positive versus negative ε-shrinkage, by the ratio of the medians for –2LL for training versus testing data, or by some other statistic and use this as a constraint in the genetic algorithm.

# Conclusions

The purpose of this work was to explain some shortcomings of current practice, to introduce a better way, and to demonstrate it with real data. The main goals were two-fold: to demonstrate the utility of bootstrap cross-validation as a practical tool for pharmacometric model selection and to evaluate various summary statistics as to which of them would be the most useful in regards to BS-CV model selection. A further pharmacodynamics goal was to see if one could use only subtle statistical differences to distinguish between different indirect response models that otherwise had very similar time course profiles (as may be useful if one were ignorant of the mechanism of action of some drug). BS-CV was applied to 13 PK models and 12 PD models using the warfarin data set often used for teaching purposes in population PK-PD and included in Monolix. BS-CV was shown to provide an objective, nonparametric basis for PK-PD model selection based on predictive ability. The most persuasive summary statistic is the $-\log(1 - r^2)$ metric for the line of observations versus individual predictions, also denoted SMPQ. This a simple, direct measure of goodness of fit that increases to infinity as residuals go to 0. For testing data, it is a direct measure of predictive accuracy.

# Acknowledgements

This work comes from my masters project, *Using Resampling to Improve Model Selection in Pharmacometrics*, defended on 2019-01-18 and chaired by Jill Fiedler-Kelly, VP of Pharmacometric Services, Cognigen Corp., a SimulationsPlus Company, and Adjunct Associate Professor, Dept. of Pharmaceutical Sciences, The University at Buffalo, SUNY, Buffalo, USA. I thank Jill for serving in this capacity. Jill was especially helpful in an earlier model selection project, which, unfortunately, did not work out as a useful methodology. Had that project been the subject of this paper, she would most definitely be listed as a co-author. I also thank Robert R. Bies, Associate Professor in the Dept. of Pharmaceutical Sciences, The University at Buffalo, for serving as my committee member and for suggesting investigating ε-shrinkage. I thank them both for their editorial comments in the earlier write-up of this work as a thesis. I also thank Greg Warnes, currently Principal Data Scientist at American Tire Distributors, for helpful conversations and for bringing to my attention the 0.632 bootstrap.

| Model | Observation model | Individual model |
|---|---|---|
| 1 | y1 = Cc + (a1 + b1×Cc) × e | log(ka) = log(ka_pop) + eta_ka |
| | *basic 1-cmt model* | log(V) = log(V_pop) + eta_V |
| | | log(Cl) = log(Cl_pop) + eta_Cl |
| 2 | y1 = Cc + (a1 + b1×Cc) × e | log(Tlag) = log(Tlag_pop) + eta_Tlag |
| | *added Tlag* | log(ka) = log(ka_pop) + eta_ka |
| | | log(V) = log(V_pop) + eta_V |
| | | log(Cl) = log(Cl_pop) + eta_Cl |
| 3 | y1 = Cc + (a1 + b1×Cc) × e | log(Tlag) = log(Tlag_pop) + eta_Tlag |
| | *added V(LnWt70)* | log(ka) = log(ka_pop) + eta_ka |
| | | log(V) = log(V_pop) + beta_V_LnWt70×LnWt70 + eta_V |
| | | log(Cl) = log(Cl_pop) + eta_Cl |
| 4 | y1 = Cc + (a1 + b1×Cc) × e | log(Tlag) = log(Tlag_pop) + eta_Tlag |
| | *added Cl(LnWt70)* | log(ka) = log(ka_pop) + eta_ka |
| | | log(V) = log(V_pop) + beta_V_LnWt70×LnWt70 + eta_V |
| | | log(Cl) = log(Cl_pop) + beta_Cl_LnWt70×LnWt70 + eta_Cl |
| 5 | y1 = Cc + (a1 + b1×Cc) × e | log(Tlag) = log(Tlag_pop) + eta_Tlag |
| | *added corr(Cl, V)* | log(ka) = log(ka_pop) + eta_ka |
| | | log(V) = log(V_pop) + beta_V_LnWt70×LnWt70 + eta_V |
| | | log(Cl) = log(Cl_pop) + beta_Cl_LnWt70×LnWt70 + eta_Cl |
| | | Correlations |
| | | id : {Cl, V} |
| 6 | y1 = Cc + (a1 + b1×Cc) × e | log(Tlag) = log(Tlag_pop) + eta_Tlag |
| | *similar to 4, but 2-cmt* | log(ka) = log(ka_pop) + eta_ka |
| | | log(Cl) = log(Cl_pop) + beta_Cl_LnWt70×LnWt70 + eta_Cl |
| | | log(V1) = log(V1_pop) + beta_V1_LnWt70×LnWt70 + eta_V1 |
| | | log(Q) = log(Q_pop) + eta_Q |
| | | log(V2) = log(V2_pop) + beta_V2_LnWt70×LnWt70 + eta_V2 |
| 7 | y1 = Cc + (a1 + b1×Cc^c1) × e | log(Tlag) = log(Tlag_pop) + eta_Tlag |
| | *changed error model (c1≠1)* | log(ka) = log(ka_pop) + eta_ka |
| | | log(Cl) = log(Cl_pop) + beta_Cl_LnWt70×LnWt70 + eta_Cl |
| | | log(V1) = log(V1_pop) + beta_V1_LnWt70×LnWt70 + eta_V1 |
| | | log(Q) = log(Q_pop) + eta_Q |
| | | log(V2) = log(V2_pop) + beta_V2_LnWt70×LnWt70 + eta_V2 |

| Model | Observation model | Individual model |
|---|---|---|
| 8 | y1 = Cc + sqrt(a1^2 + (b1×Cc^c1)^2) × e | log(Tlag) = log(Tlag_pop) + eta_Tlag |
| *changed error model to (full) combined2 (c1≠1)* | | log(ka) = log(ka_pop) + eta_ka |
| | | log(Cl) = log(Cl_pop) + beta_Cl_LnWt70×LnWt70 + eta_Cl |
| | | log(V1) = log(V1_pop) + beta_V1_LnWt70×LnWt70 + eta_V1 |
| | | log(Q) = log(Q_pop) + eta_Q |
| | | log(V2) = log(V2_pop) + beta_V2_LnWt70×LnWt70 + eta_V2 |
| 9 | y1 = Cc + sqrt(a1^2 + (b1×Cc)^2) × e | log(Tlag) = log(Tlag_pop) + eta_Tlag |
| *simplified combined2 error model* | | log(ka) = log(ka_pop) + eta_ka |
| | | log(Cl) = log(Cl_pop) + beta_Cl_LnWt70×LnWt70 + eta_Cl |
| | | log(V1) = log(V1_pop) + beta_V1_LnWt70×LnWt70 + eta_V1 |
| | | log(Q) = log(Q_pop) + eta_Q |
| | | log(V2) = log(V2_pop) + beta_V2_LnWt70×LnWt70 + eta_V2 |
| 10 | y1 = Cc + (a1 + b1×Cc) × e | log(Tlag) = log(Tlag_pop) + eta_Tlag |
| *Cl(LnAge31, LnWt70), combined1 error model* | | log(ka) = log(ka_pop) + eta_ka |
| | | log(Cl) = log(Cl_pop)+ beta_Cl_LnAge31×LnAge31 + beta_Cl_LnWt70×LnWt70 + eta_Cl |
| | | log(V1) = log(V1_pop) + beta_V1_LnWt70×LnWt70 + eta_V1 |
| | | log(Q) = log(Q_pop) + eta_Q |
| | | log(V2) = log(V2_pop) + beta_V2_LnWt70×LnWt70 + eta_V2 |
| 11 | y1 = Cc + sqrt(a1^2 + (b1×Cc)^2) × e | log(Tlag) = log(Tlag_pop) + eta_Tlag |
| *Cl(LnAge31, LnWt70) combined2 error model* | | log(ka) = log(ka_pop) + eta_ka |
| | | log(Cl) = log(Cl_pop) + beta_Cl_LnAge31×LnAge31 + beta_Cl_LnWt70×LnWt70 + eta_Cl |
| | | log(V1) = log(V1_pop) + beta_V1_LnWt70×LnWt70 + eta_V1 |
| | | log(Q) = log(Q_pop) + beta_Q_LnAge31×LnAge31 + eta_Q |
| | | log(V2) = log(V2_pop) + beta_V2_LnWt70×LnWt70 + eta_V2 |
| 12 | y1 = Cc + (a1 + b1×Cc) × e | log(Tlag) = log(Tlag_pop) + eta_Tlag |
| | | log(ka) = log(ka_pop) + eta_ka |

| Model | Observation model | Individual model |
|---|---|---|
| 13 | $y1 = Cc + \sqrt{a1^2 + (b1 \times Cc)^2} \times e$ | *Cl(LnAge31), combined1 error model* |
| | | log(Cl) = log(Cl_pop) + beta_Cl_LnAge31×LnAge31 + eta_Cl |
| | | log(V1) = log(V1_pop) + beta_V1_LnWt70×LnWt70 + eta_V1 |
| | | log(Q) = log(Q_pop) + eta_Q |
| | | log(V2) = log(V2_pop) + beta_V2_LnWt70×LnWt70 + eta_V2 |
| | | log(Tlag) = log(Tlag_pop) + eta_Tlag |
| | | *Cl(LnAge31), combined2 error model* |
| | | log(ka) = log(ka_pop) + eta_ka |
| | | log(Cl) = log(Cl_pop) + beta_Cl_LnAge31×LnAge31 + eta_Cl |
| | | log(V1) = log(V1_pop) + beta_V1_LnWt70×LnWt70 + eta_V1 |
| | | log(Q) = log(Q_pop) + eta_Q |
| | | log(V2) = log(V2_pop) + beta_V2_LnWt70×LnWt70 + eta_V2 |

**Table 1**. Pharmacokinetic error (observation) and individual models for the 1 and 2 compartment structural models 1-6 and 7-13 respectively. The combined1 and combined2 error models are Monolix' terminology. Covariates were scaled so that LnWt70 refers to log(weight/70) where log is the natural logarithm, 70 is the mean weight (in kg) of the subjects, and likewise LnAge31 for scaled age.

| Model | Median SMPQ |
|---|---|
| 1 | 4.959724 |
| 2 | 4.950134 |
| 3 | 4.932914 |
| 4 | 4.896363 |
| 5 | 4.925444 |
| 6 | 4.943017 |
| 7 | 4.905763 |
| 8 | 4.925442 |
| 9 | 4.927192 |
| 10 | 4.881644 |
| 11 | 4.934955 |
| 12 | 4.910872 |

**Table 2.** The predictive ability of the IRM PD models evaluated is all about the same.

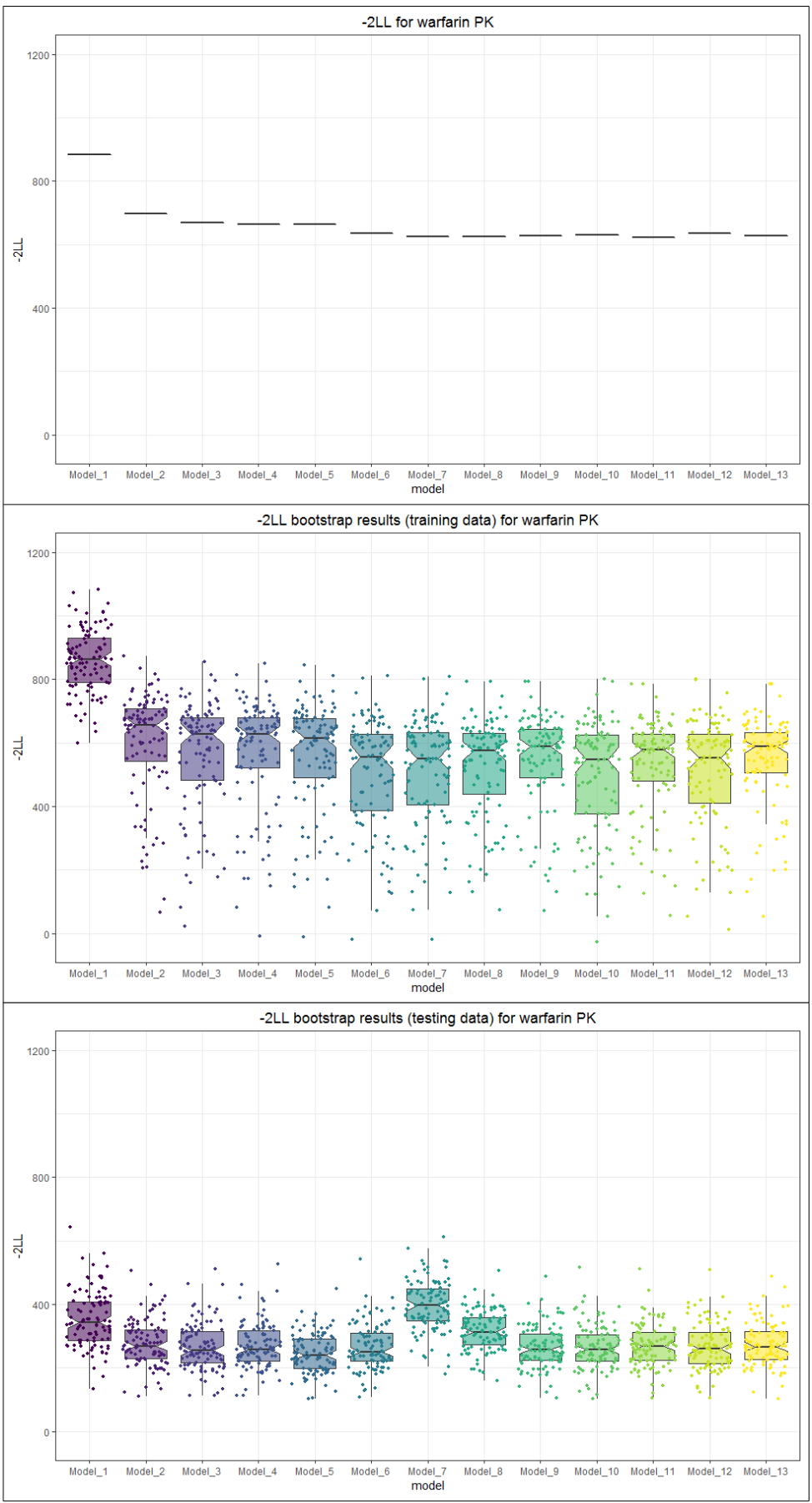

**Figure 1.** –2LL for original, training, and testing data.

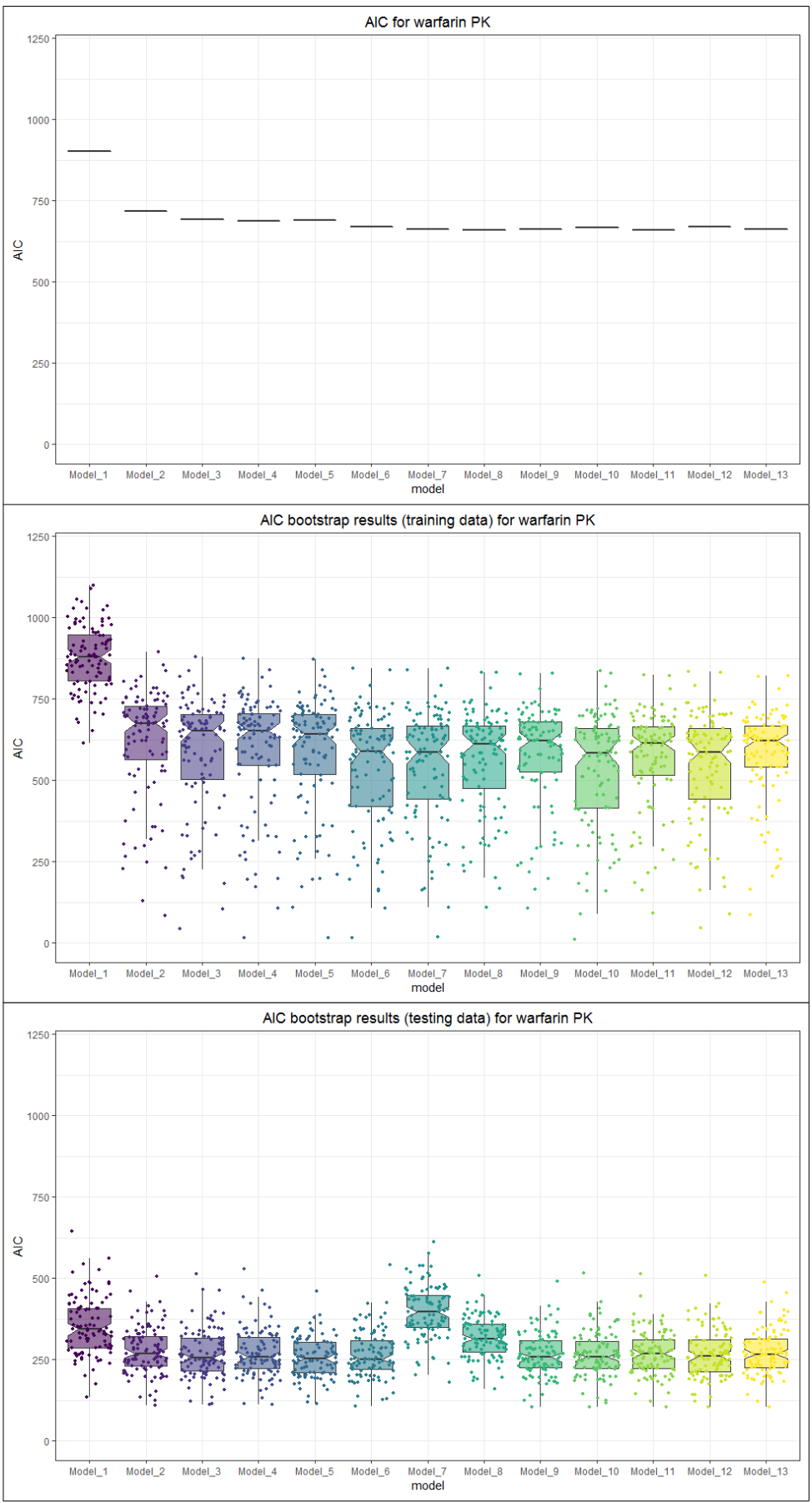

**Figure 2.** AIC for original, training, and testing data.

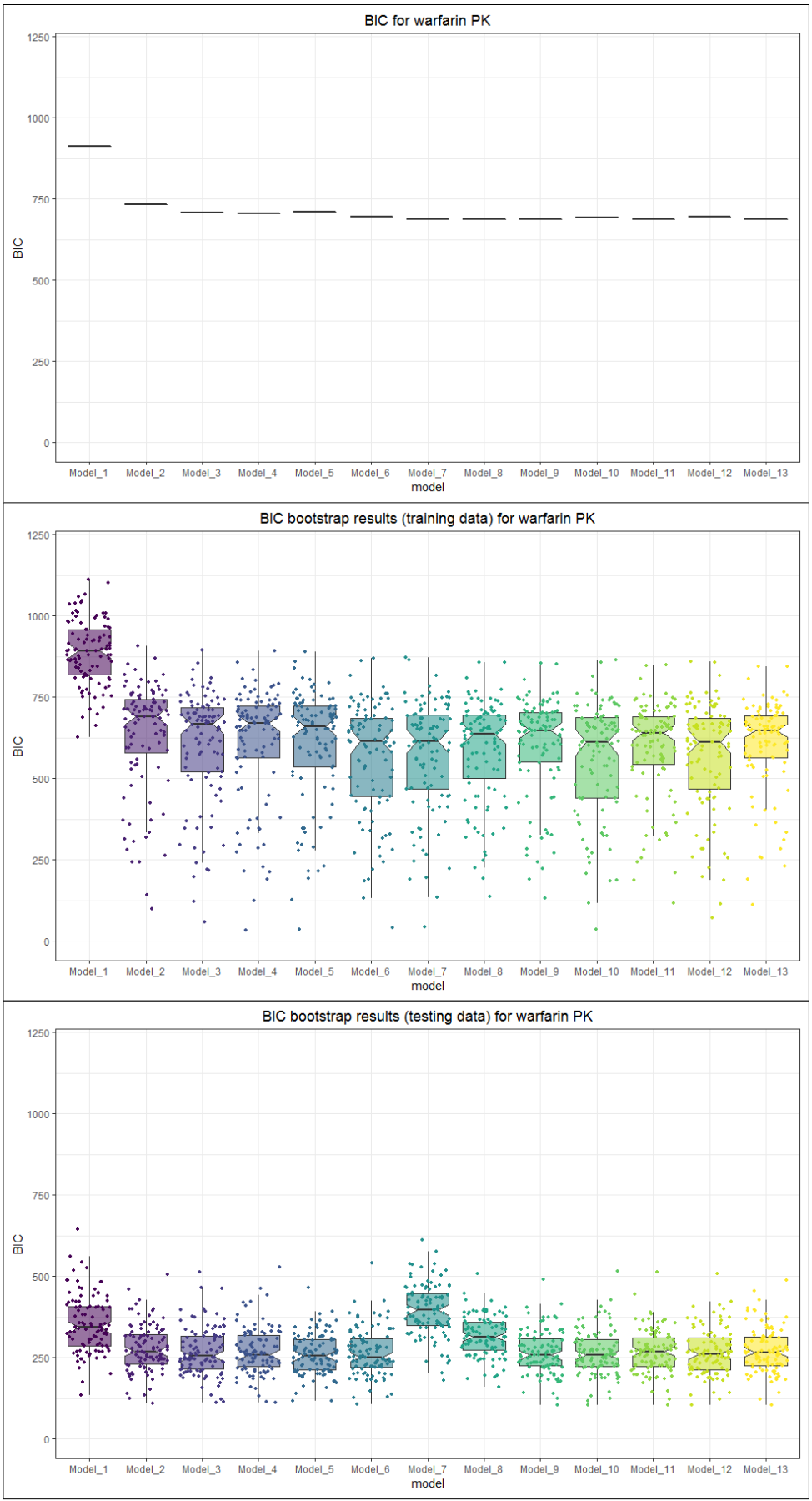

**Figure 3.** BIC for original, training, and testing data.

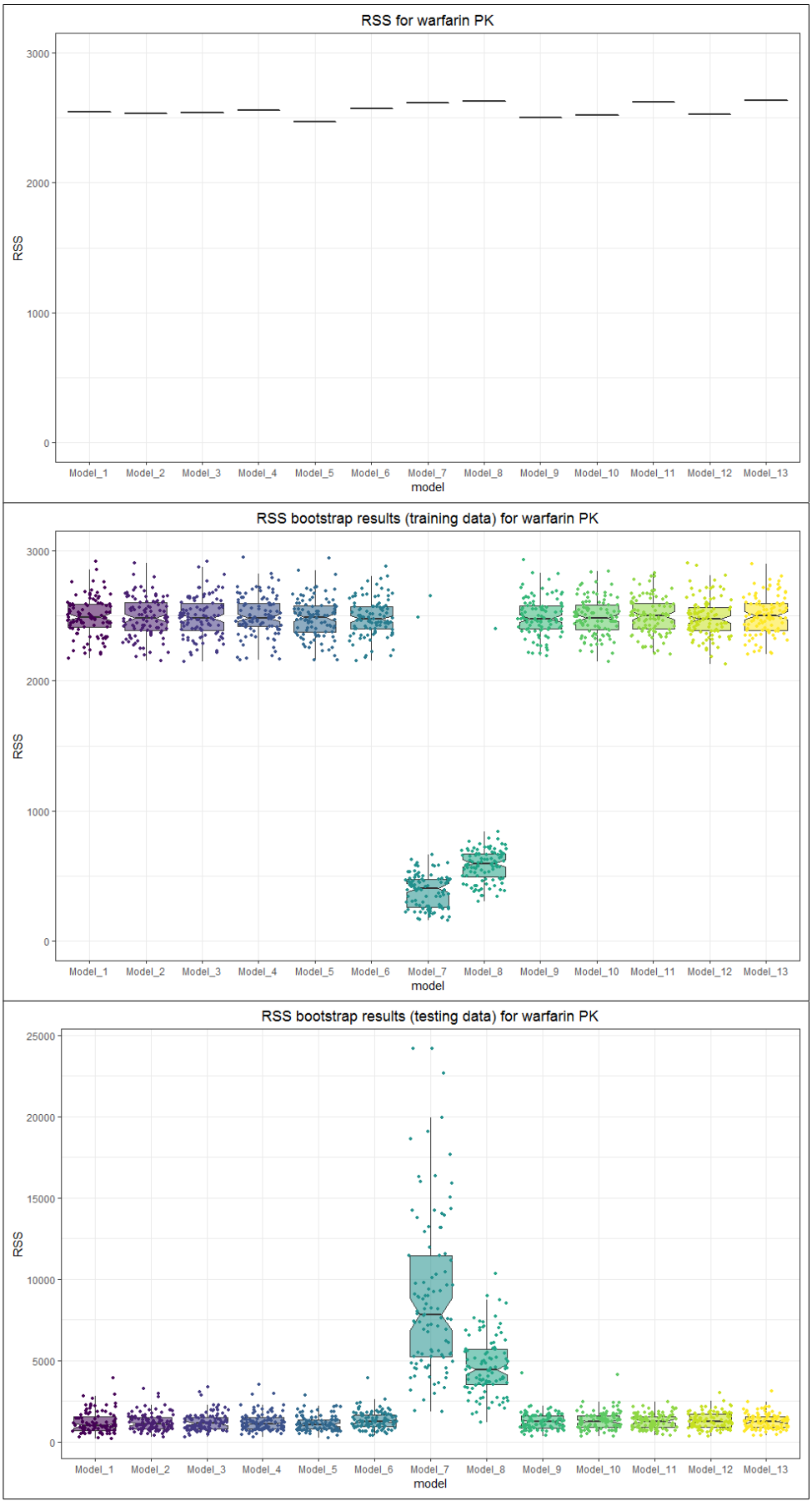

**Figure 4.** RSS for original, training, and testing data.

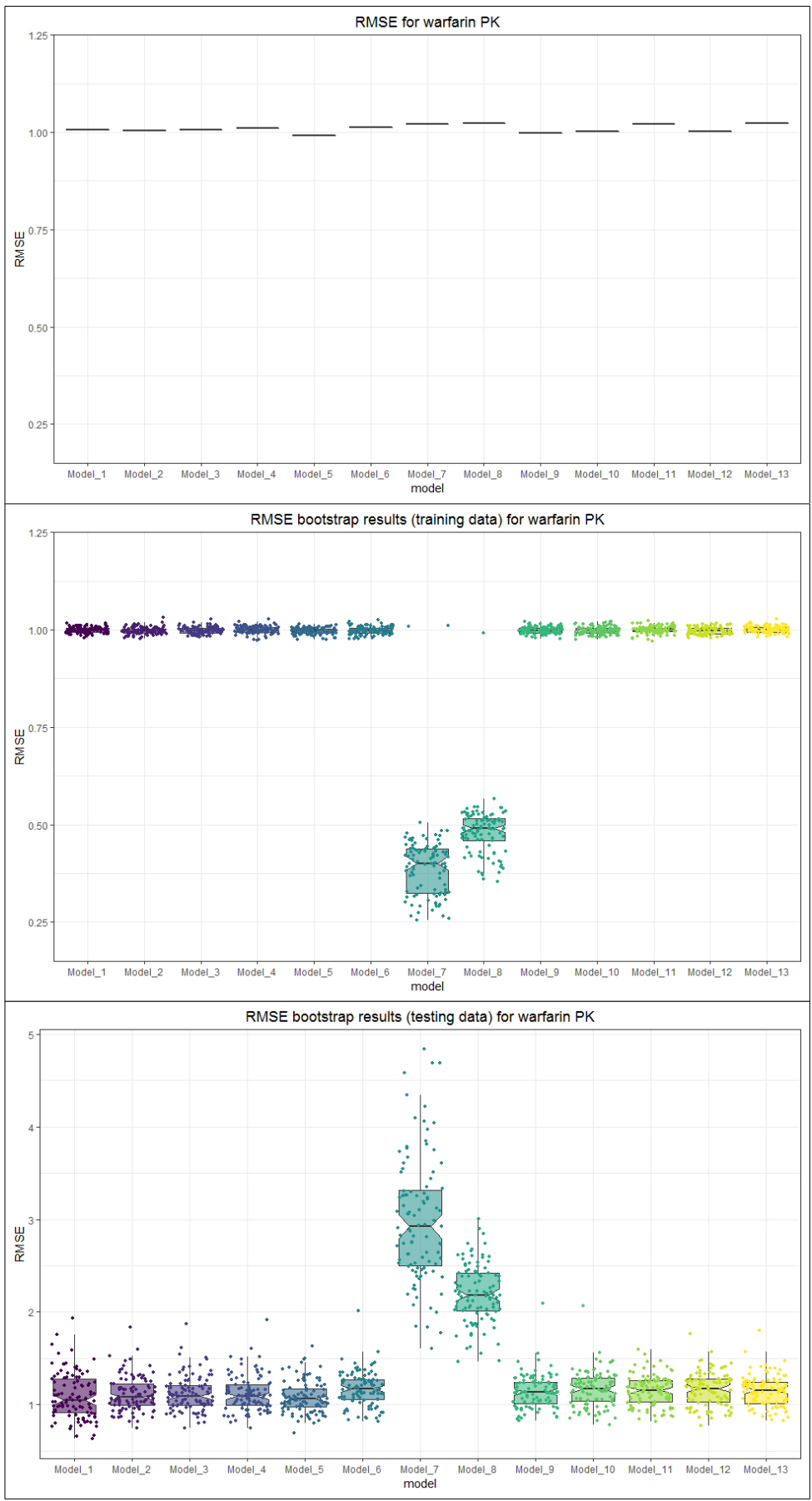

**Figure 5.** RMSE for original, training, and testing data.

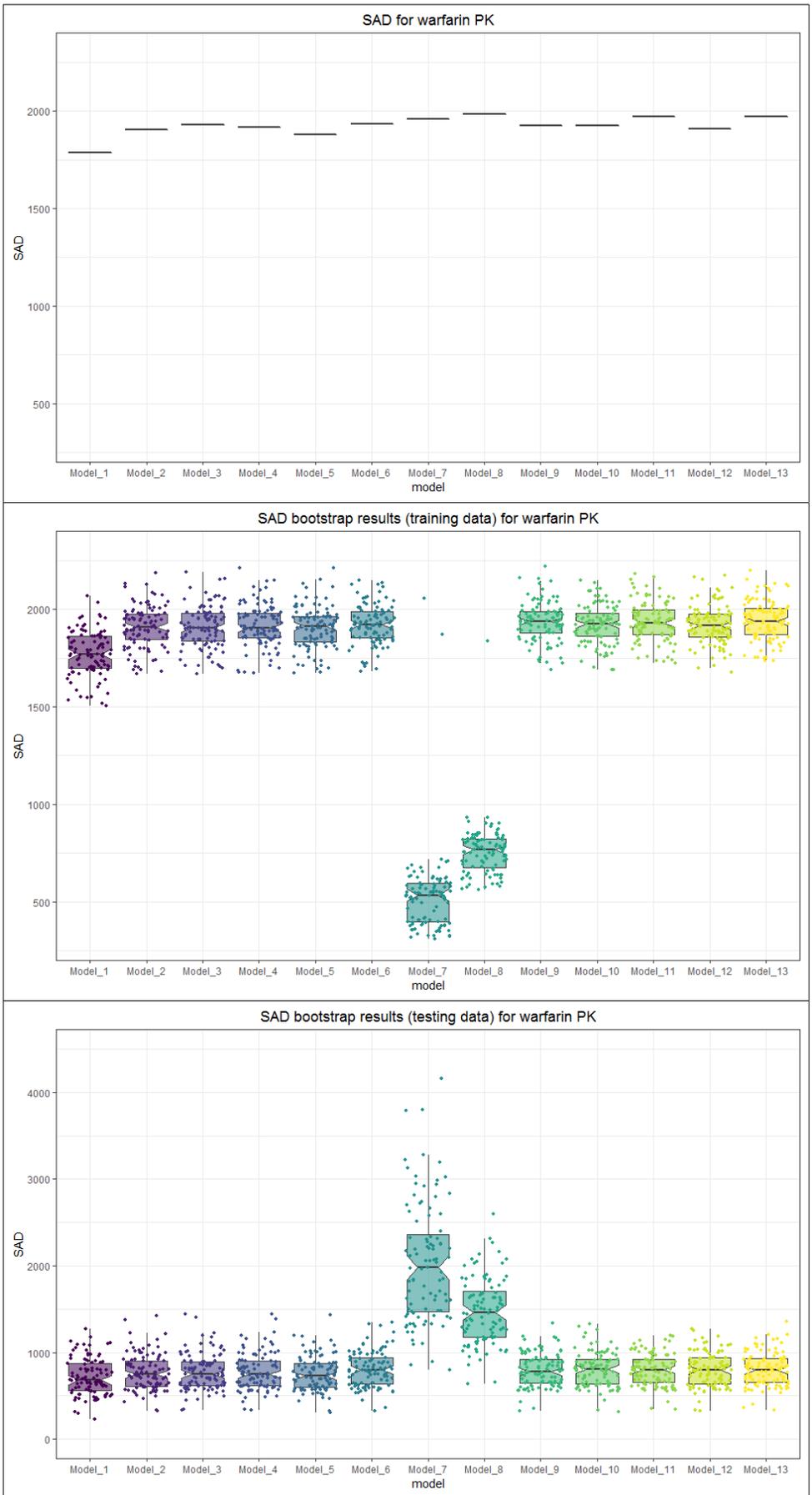

**Figure 6.** SAD for original, training, and testing data.

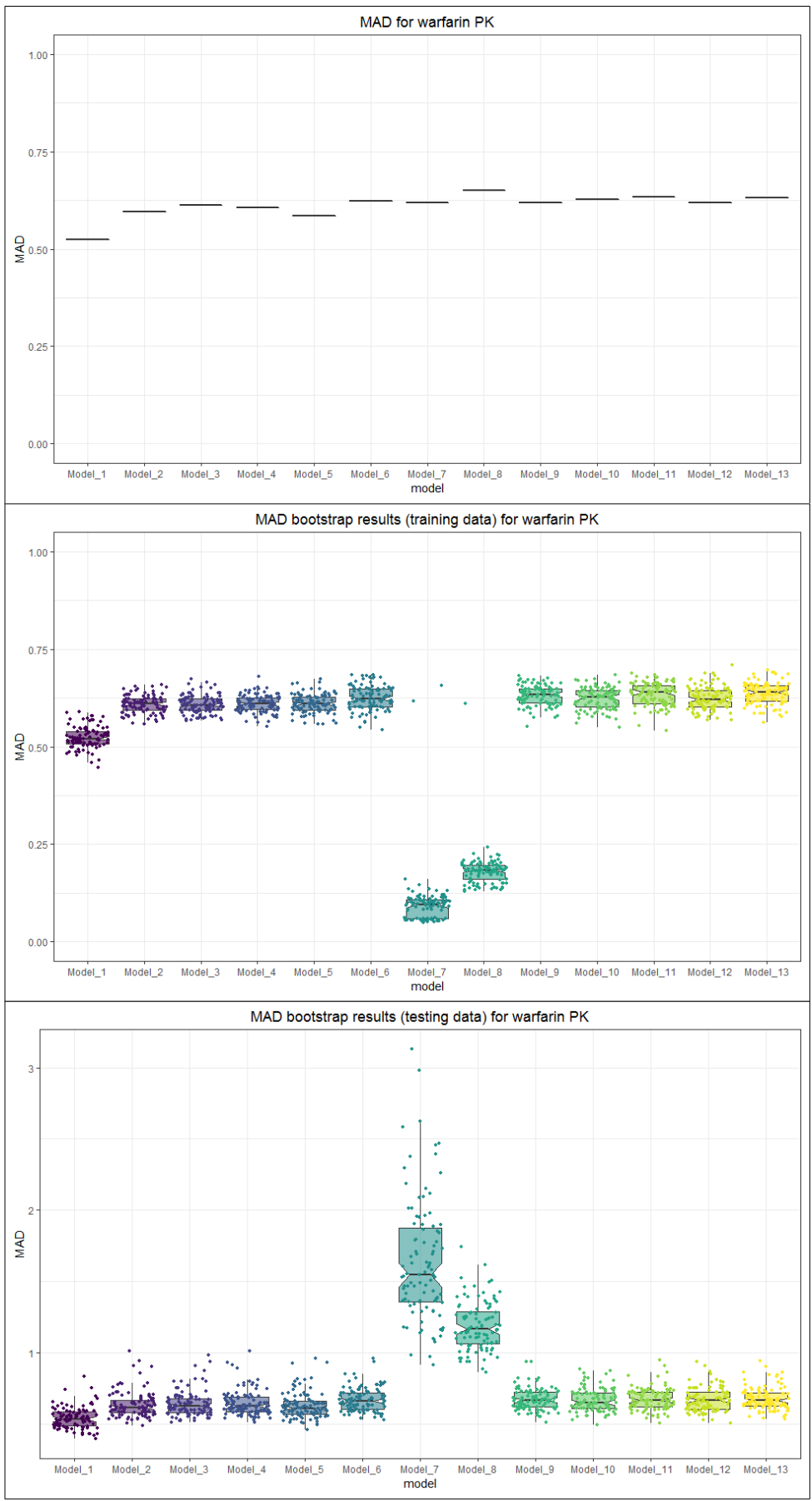

**Figure 7.** MAD for original, training, and testing data.

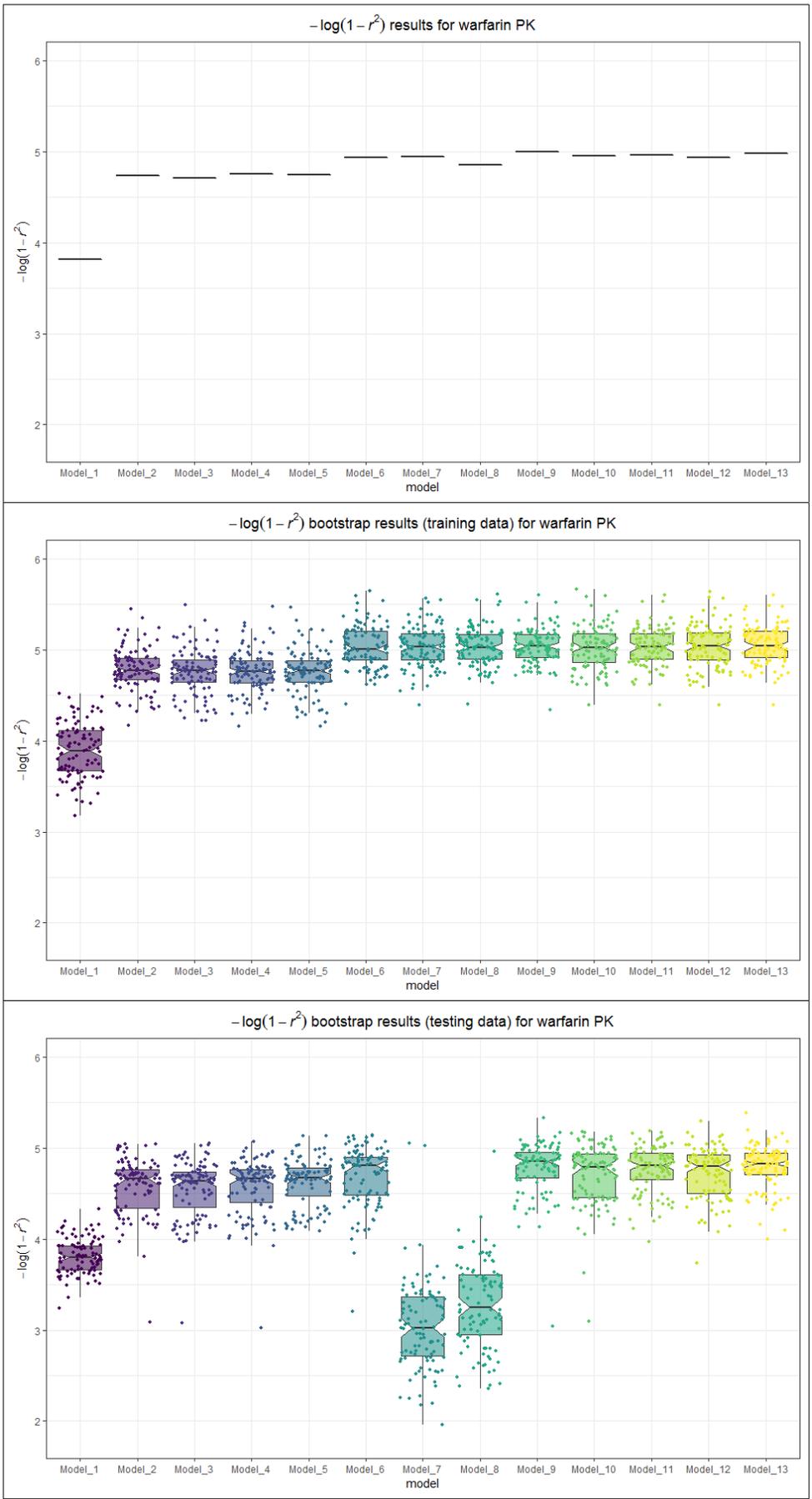

**Figure 8.** SMPQ for original, training, and testing data.

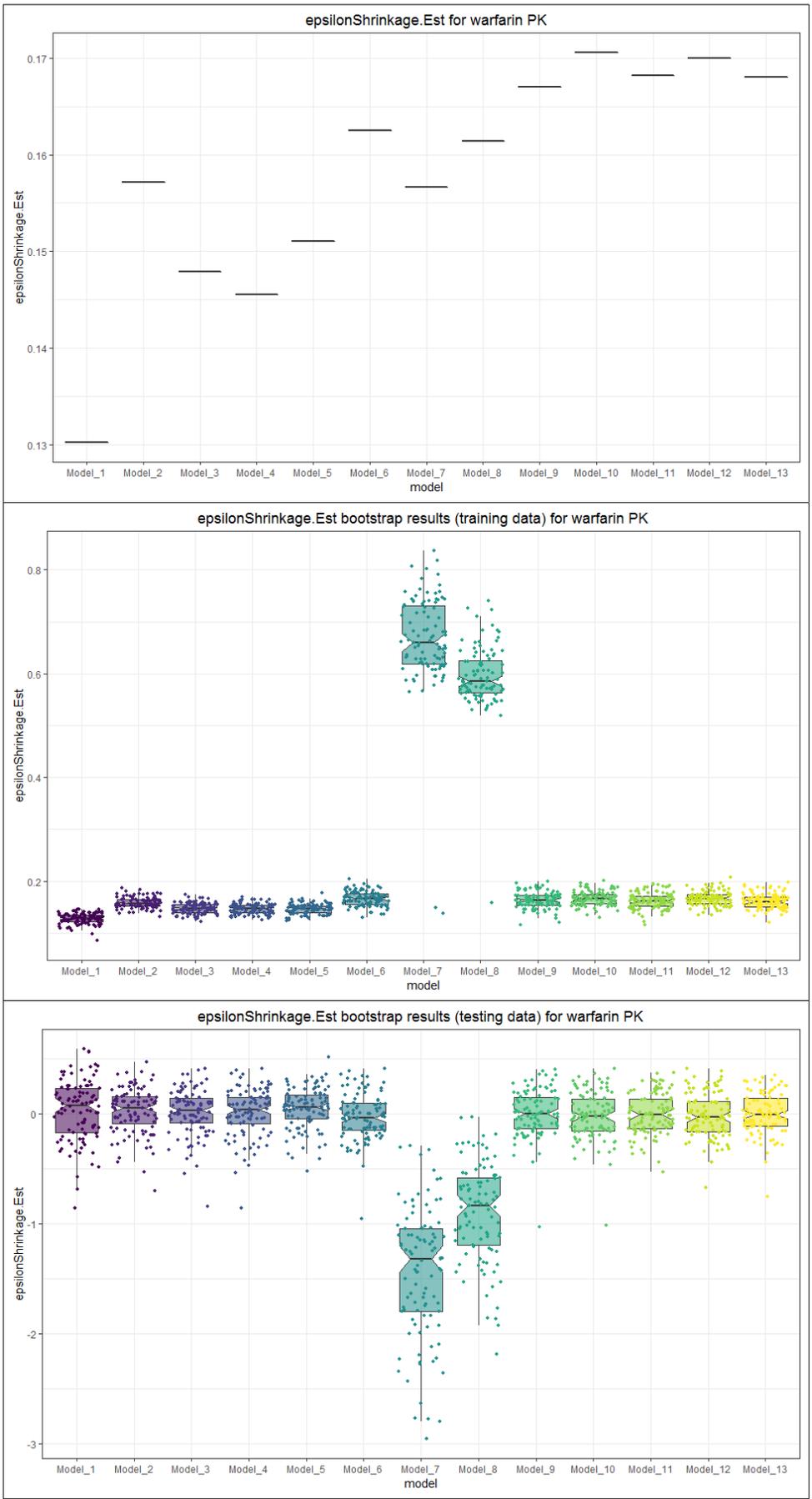

**Figure 9.** ε-Shrinkage (via EBE) for original, training, and testing data.

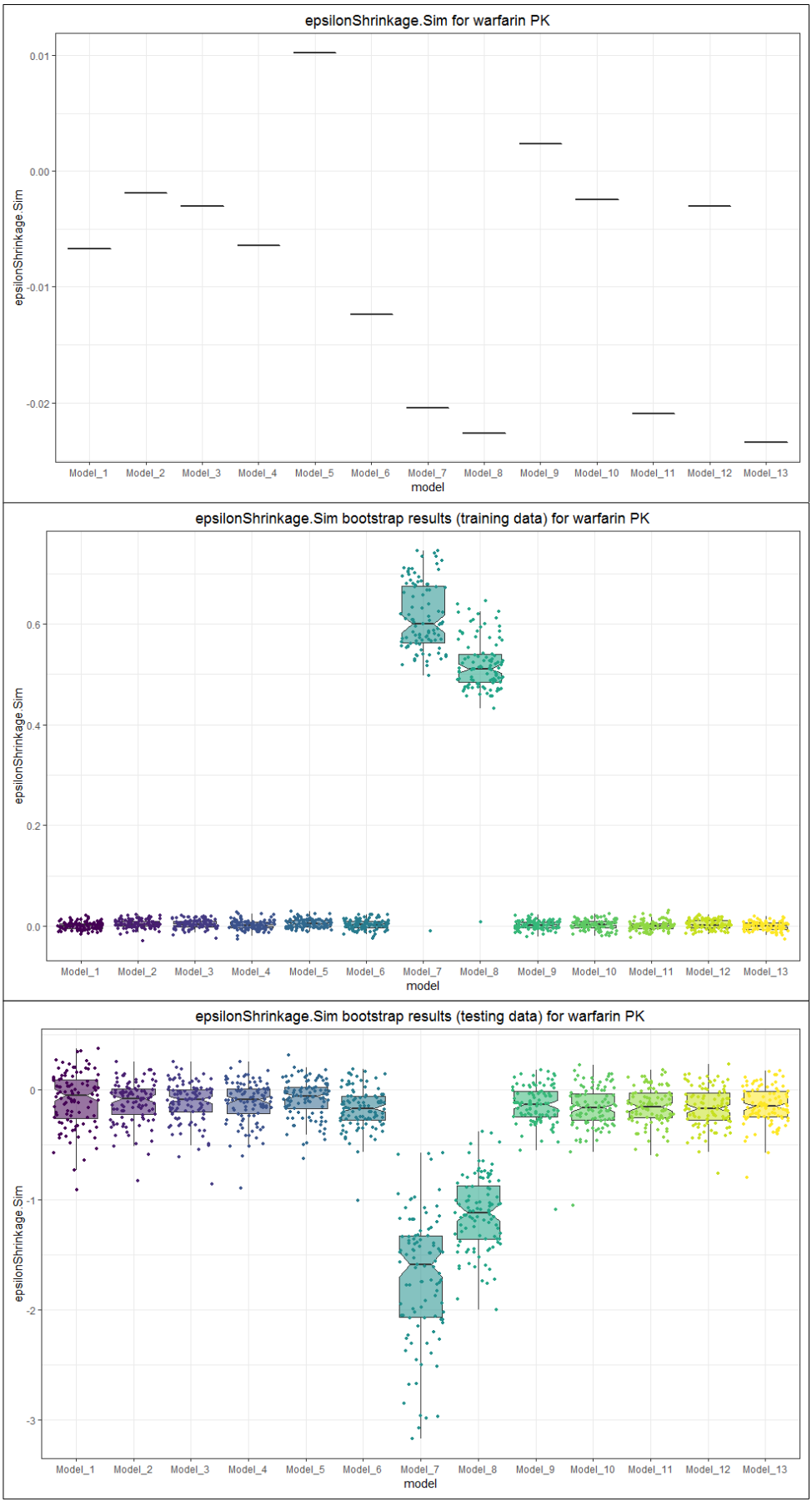

**Figure 10.** ε-Shrinkage (via simulations) for original, training, and testing data.

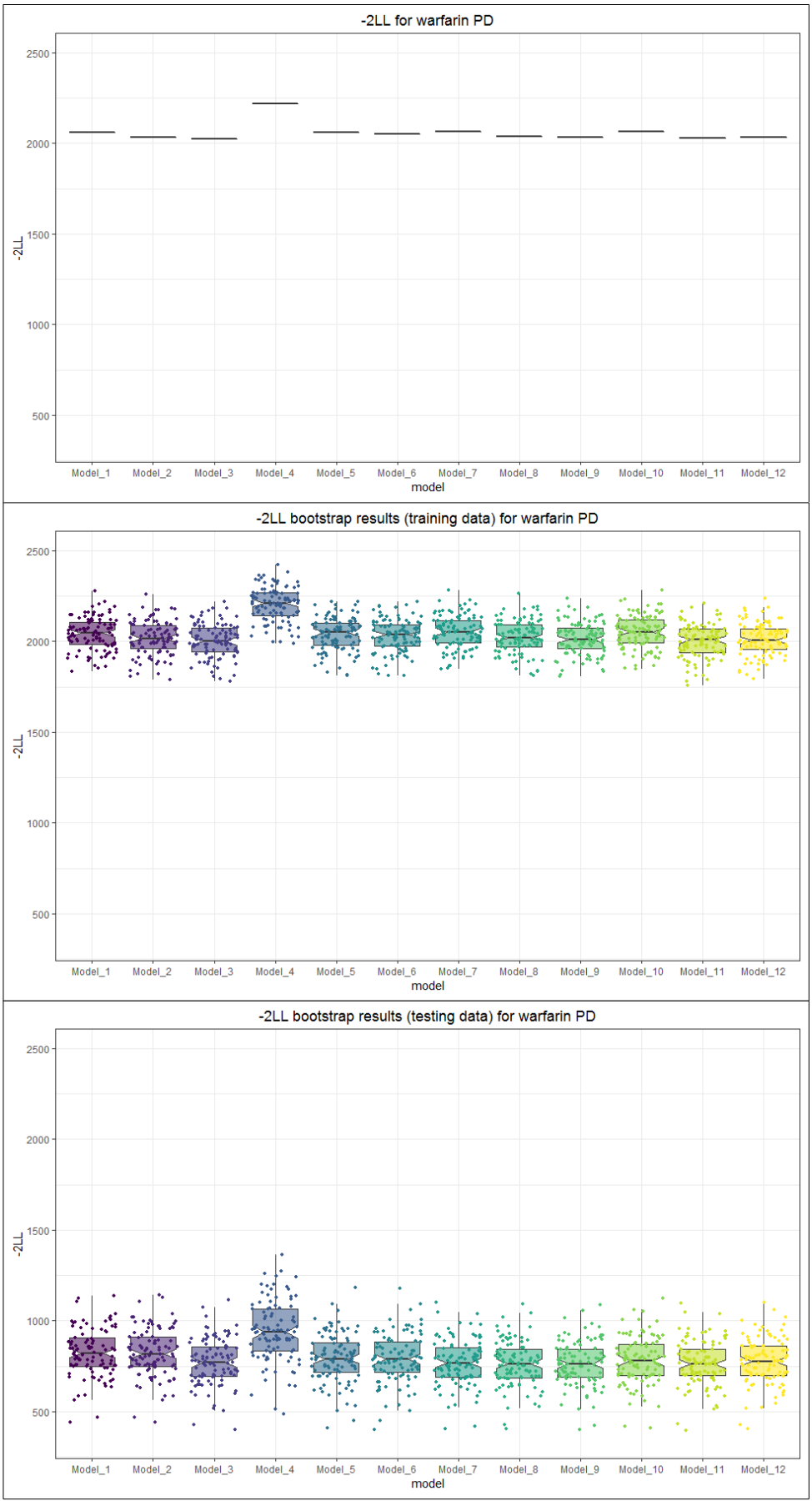

**Figure 11.** –2LL for original, training, and testing data.

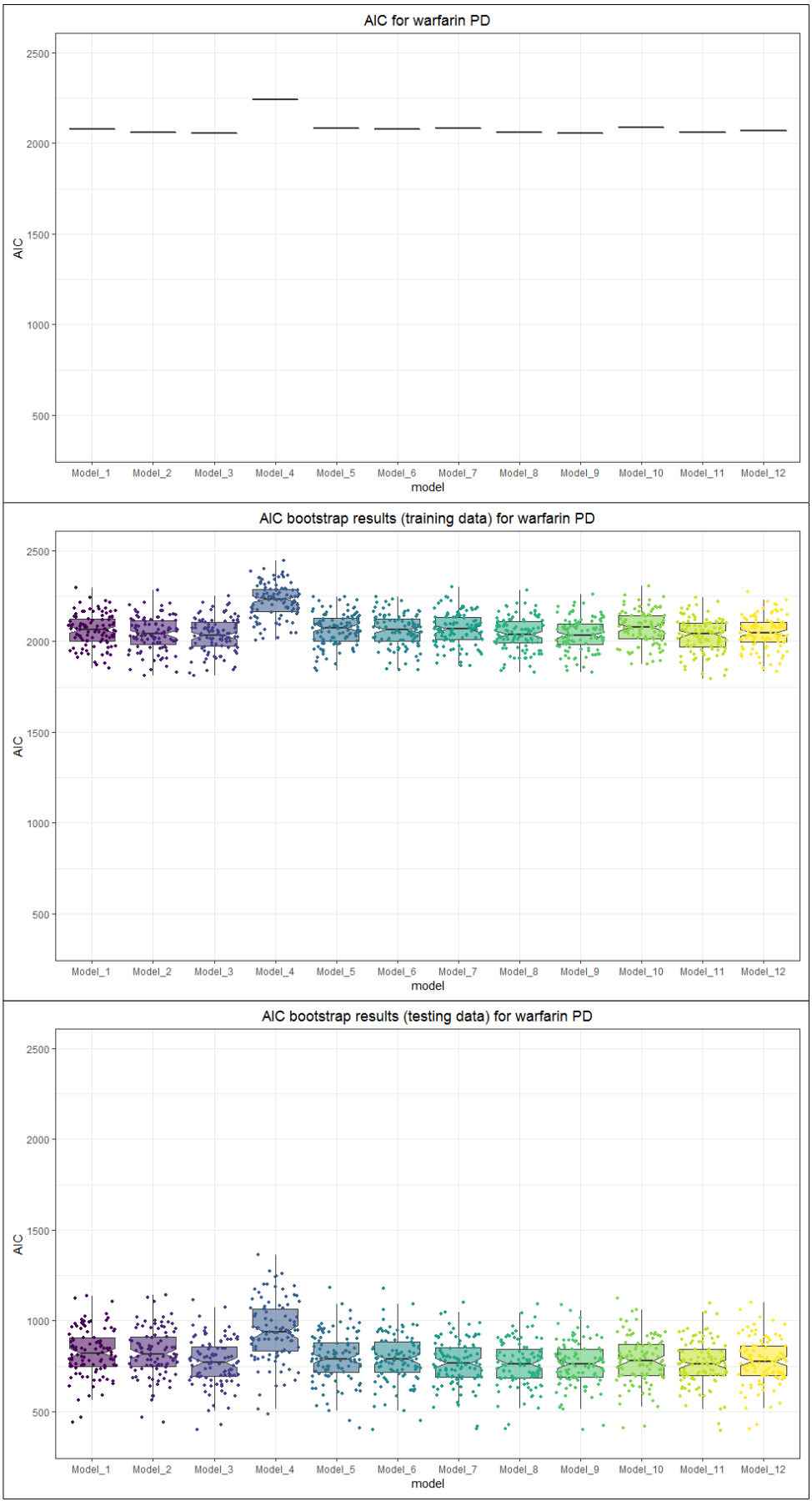

**Figure 12.** AIC for original, training, and testing data.

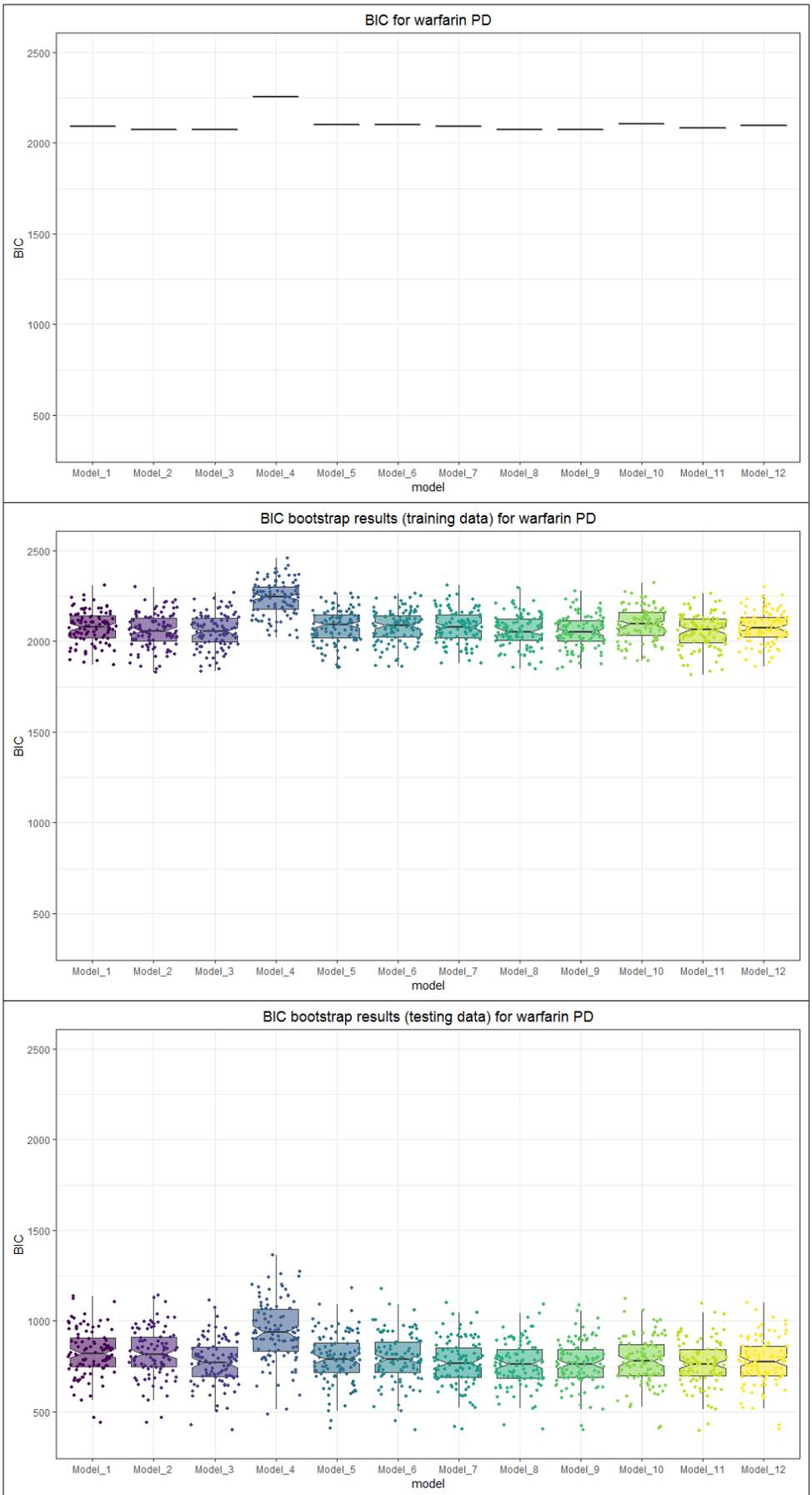

**Figure 13.** BIC for original, training, and testing data.

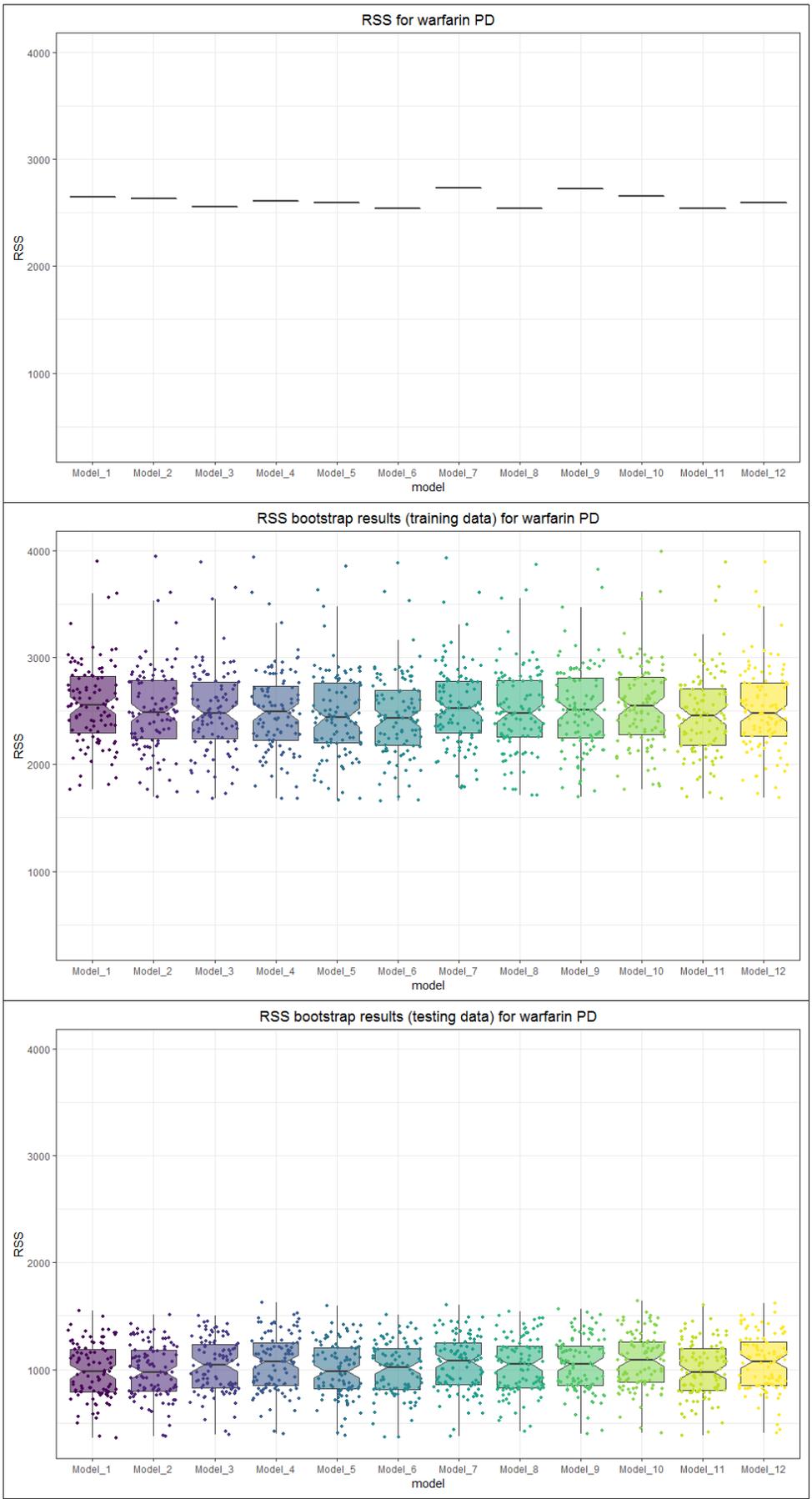

**Figure 14.** RSS for original, training, and testing data.

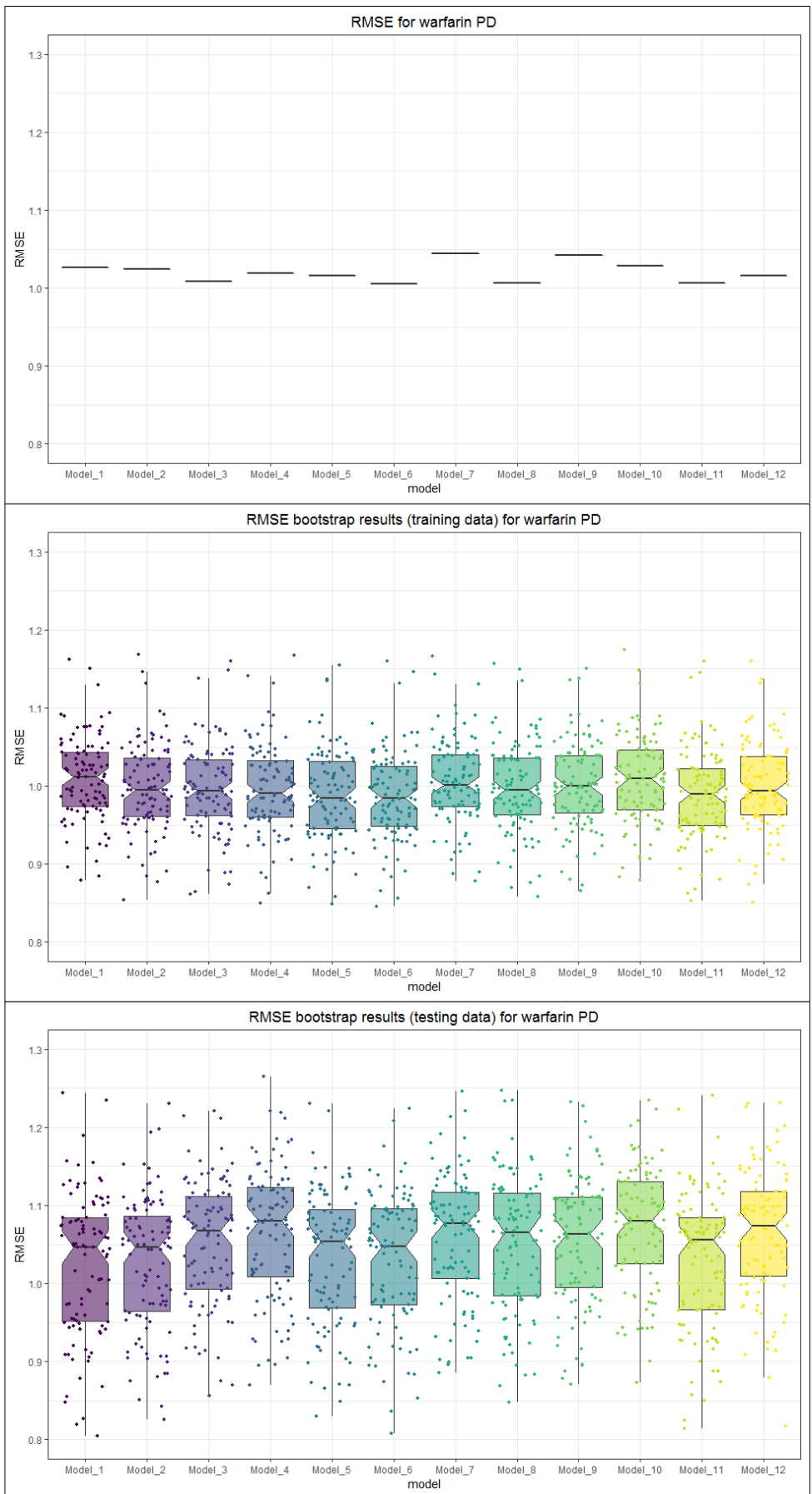

**Figure 15.** RMSE for original, training, and testing data.

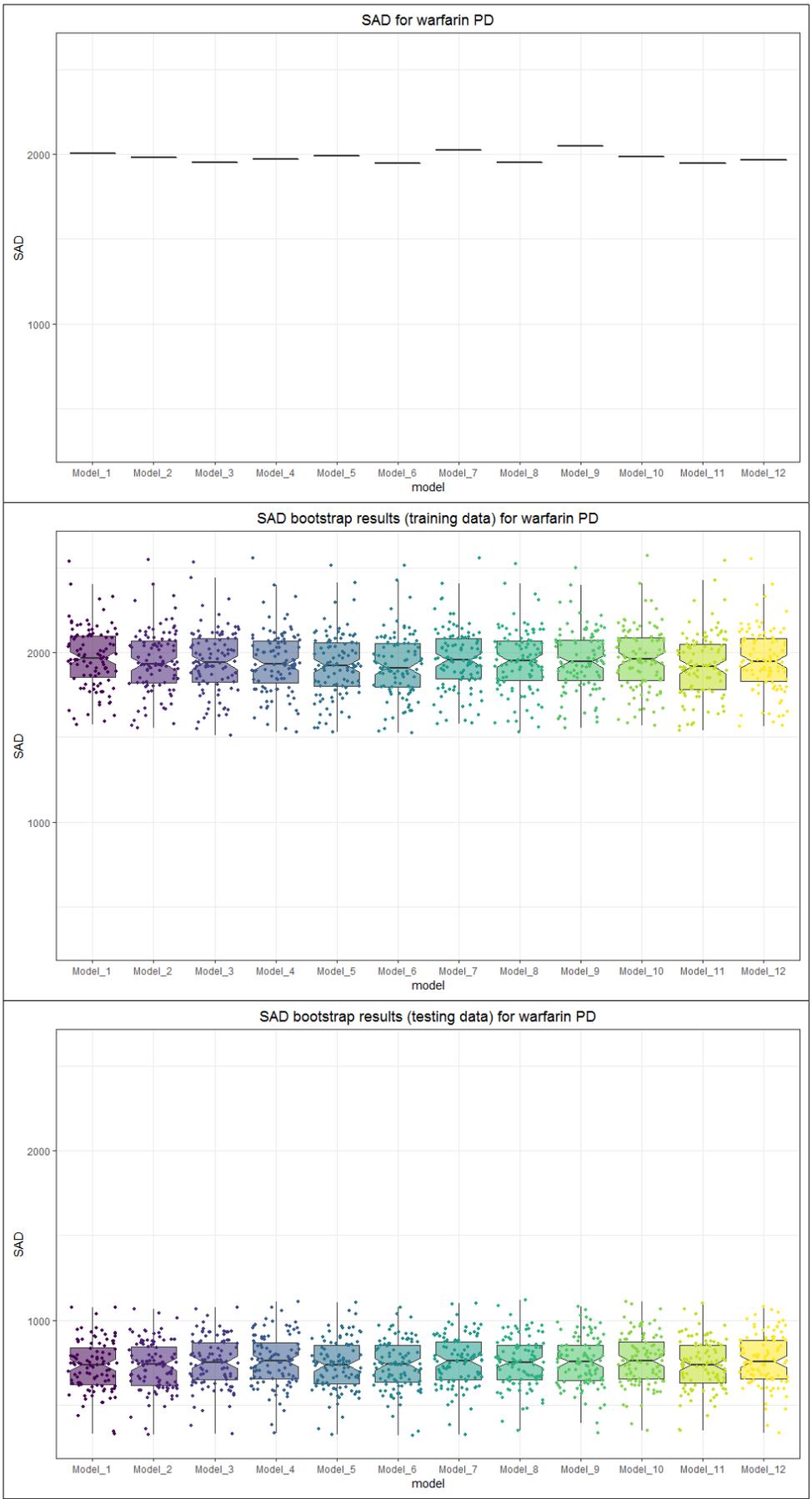

**Figure 16.** SAD for original, training, and testing data.

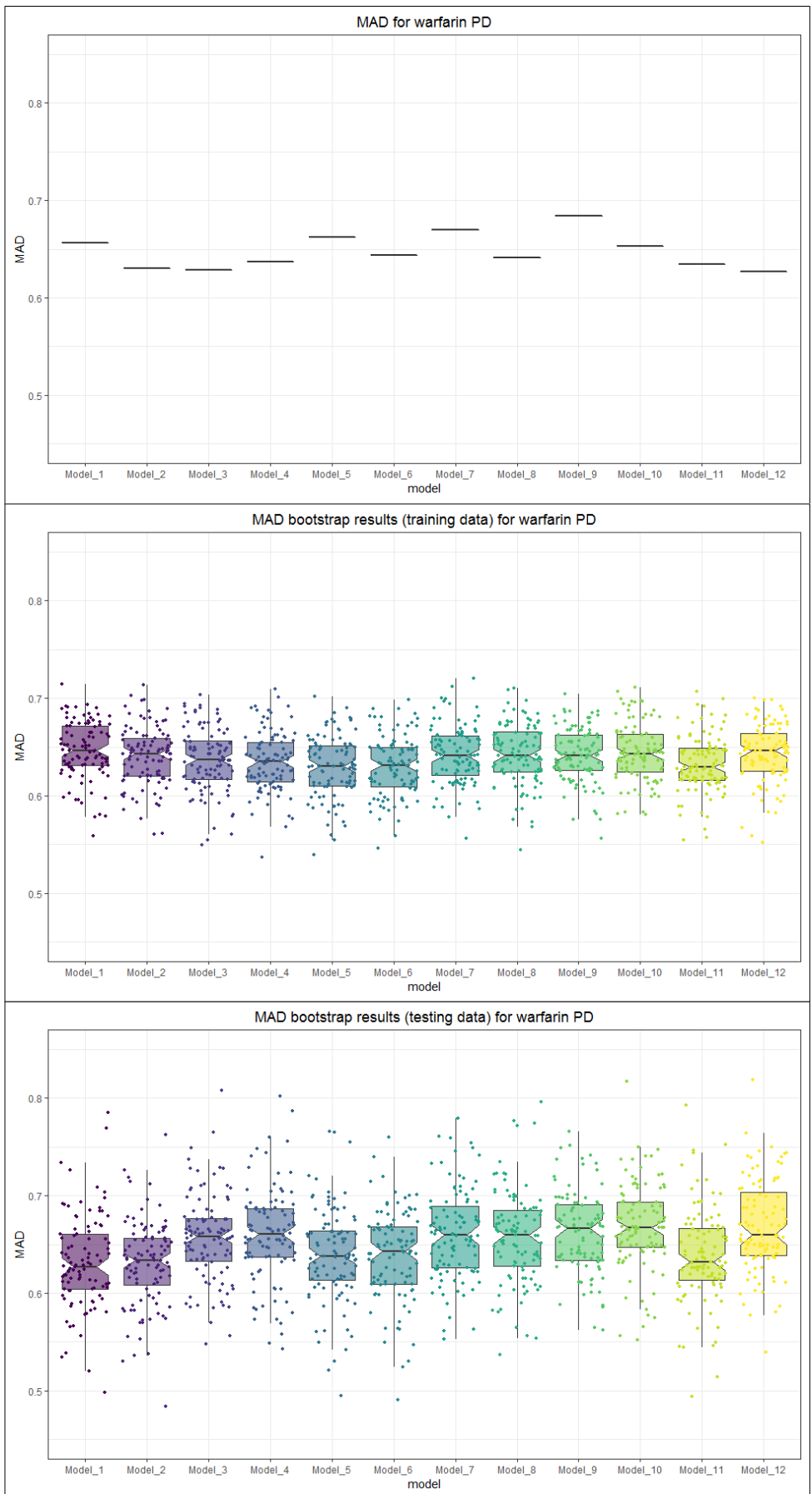

**Figure 17.** MAD for original, training, and testing data.

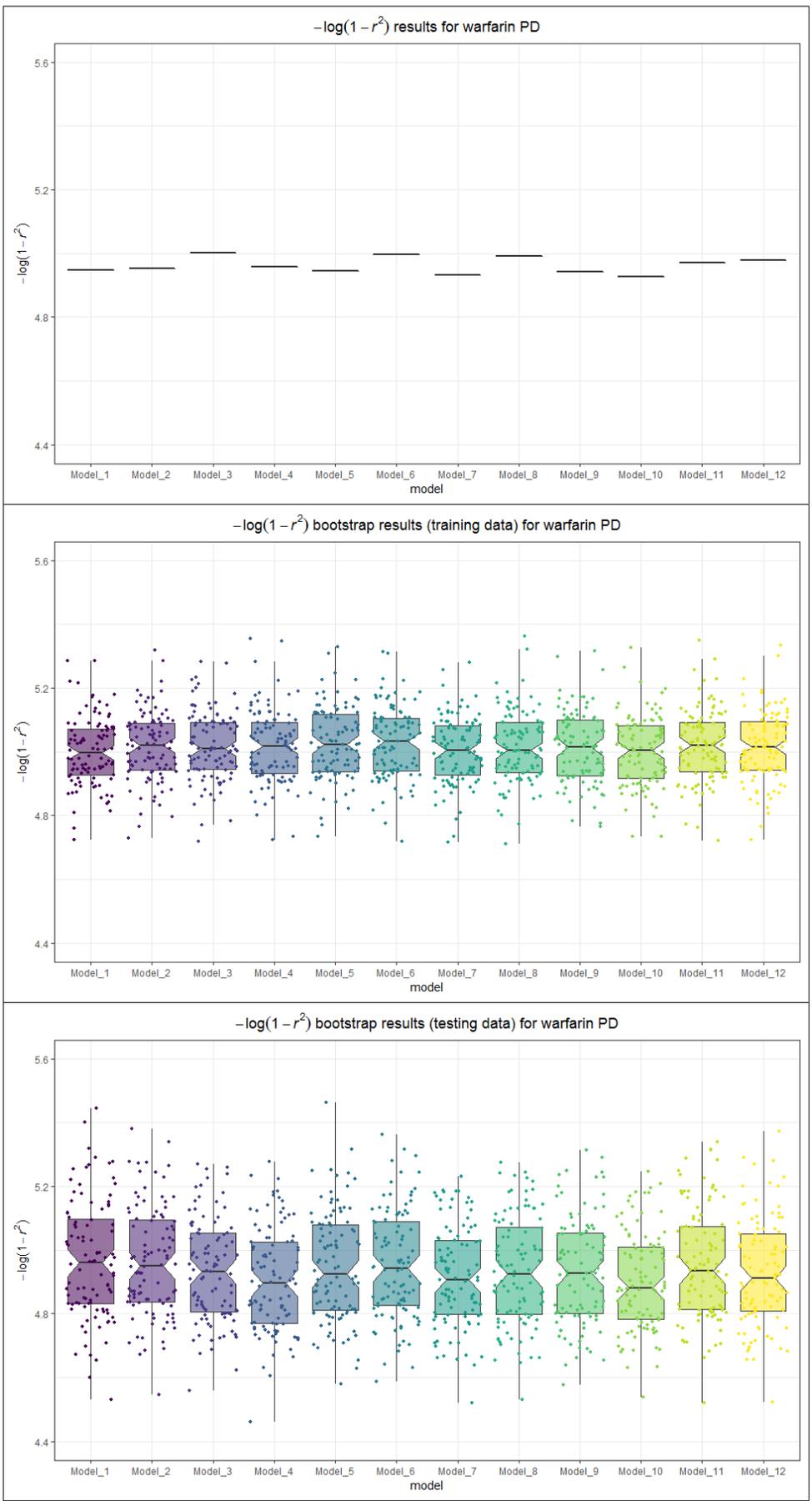

**Figure 18.** SMPQ for original, training, and testing data.

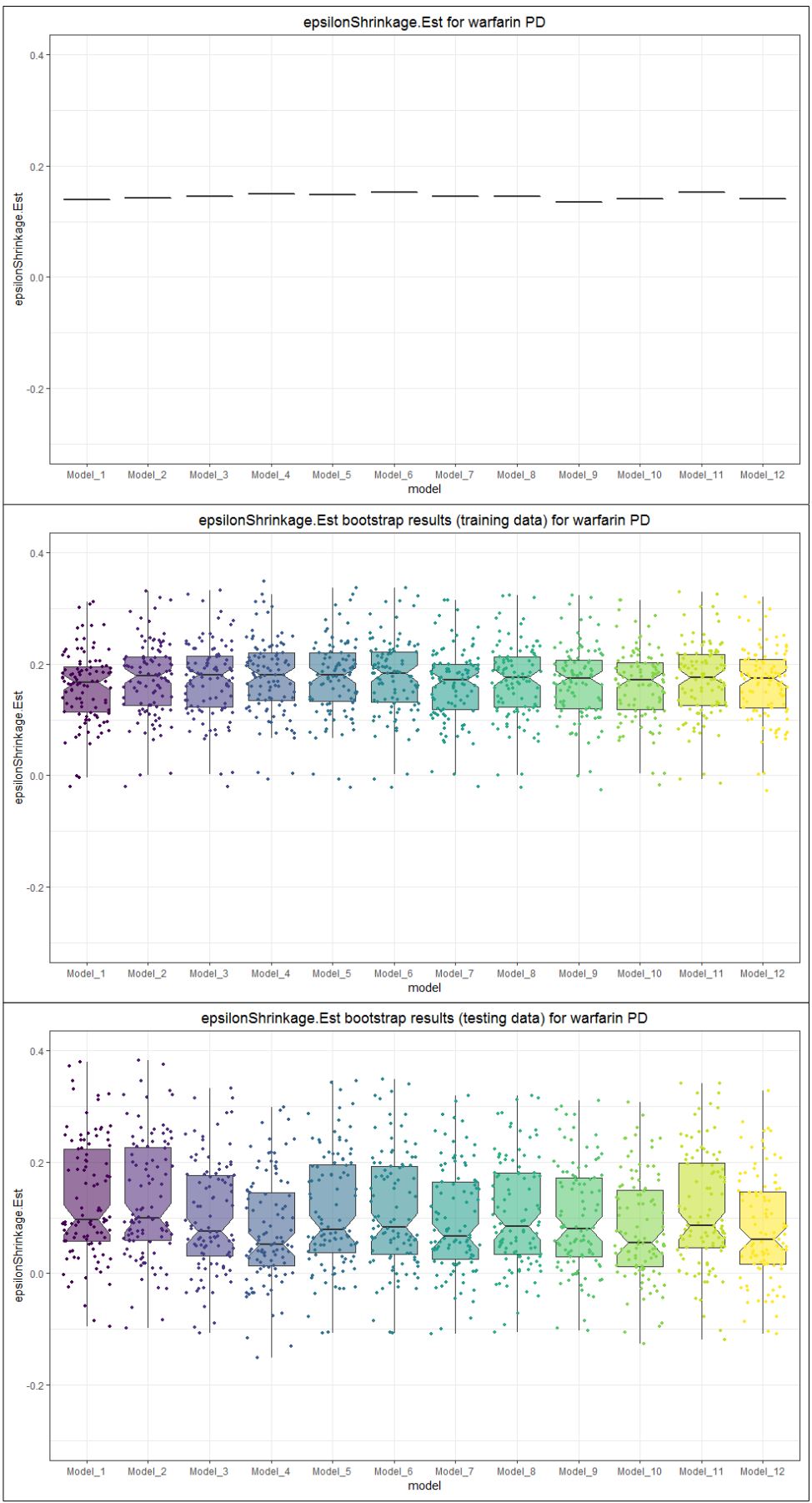

**Figure 19.** ε-Shrinkage (via EBE) for original, training, and testing data.

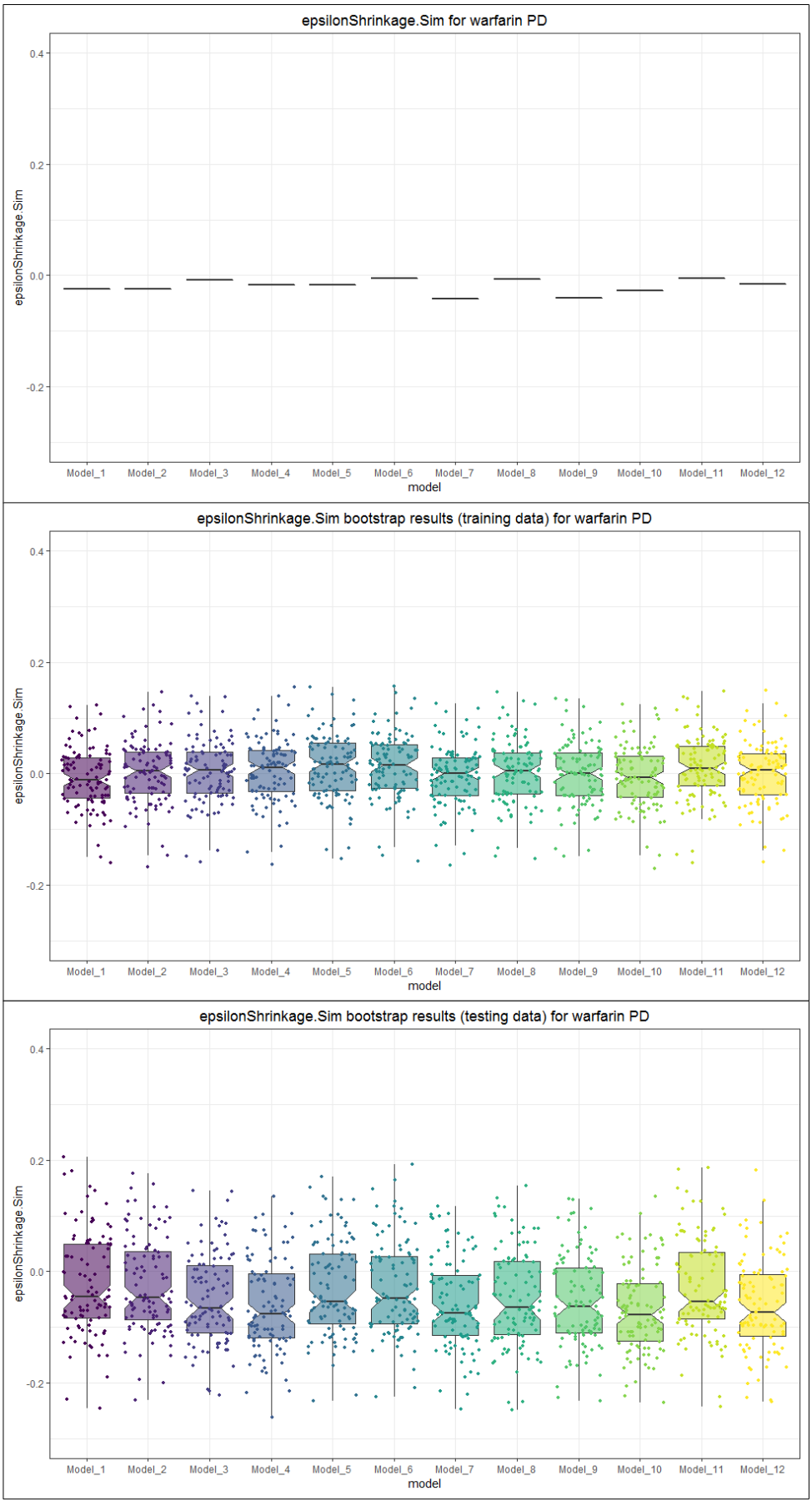

**Figure 20.** ε-Shrinkage (via simulations) for original, training, and testing data.